\documentclass[a4paper,twocolumn,11pt,accepted=2024-09-02]{quantumarticle}
\pdfoutput=1
\usepackage[utf8]{inputenc}
\usepackage[english]{babel}
\usepackage[T1]{fontenc}
\usepackage{amsthm,amsfonts,amstext,amssymb,amsmath}
\usepackage{physics}
\usepackage[ruled, lined]{algorithm2e}
\usepackage{hyperref}
\usepackage{float}
\usepackage{tikz}
\usepackage{lipsum}

\usepackage[numbers,sort&compress]{natbib}

\begin{document}

\title{Universal quantum processors in spin systems via robust local pulse sequences}

\author{Matteo Votto}
\email{matteo.votto@lpmmc.cnrs.fr}
\affiliation{Univ. Grenoble Alpes, CNRS, LPMMC, 38000 Grenoble, France}

\author{Johannes Zeiher}
\affiliation{Fakult\"at f\"ur Physik, Ludwig-Maximilians-Universit\"at M\"unchen, 80799 M\"unchen, Germany}
\affiliation{Max-Planck-Institut f\"ur Quantenoptik, 85748 Garching, Germany}	
\affiliation{Munich Center for Quantum Science and Technology (MCQST), 80799 Munich, Germany}

\author{Beno\^it Vermersch}
\affiliation{Univ. Grenoble Alpes, CNRS, LPMMC, 38000 Grenoble, France}

\begin{abstract}
    We propose a protocol to realize quantum simulation and computation in spin systems with long-range interactions. 
    Our approach relies on the local addressing of single spins with external fields parametrized by Walsh functions.
    This enables a mapping from a class of target Hamiltonians, defined by the graph structure of their interactions, to pulse sequences.
    We then obtain a recipe to implement arbitrary two-body Hamiltonians and universal quantum circuits.
    Performance guarantees are provided in terms of bounds on Trotter errors and total number of pulses.
    Additionally, Walsh pulse sequences are shown to be robust against various types of pulse errors, in contrast to previous hybrid digital-analog schemes of quantum computation.
    We demonstrate and numerically benchmark our protocol with examples from the dynamics of spin models, quantum error correction and quantum optimization algorithms.
\end{abstract}

\maketitle

\section{Introduction}

\begin{figure*}[t]
    \includegraphics[width=0.99\textwidth]{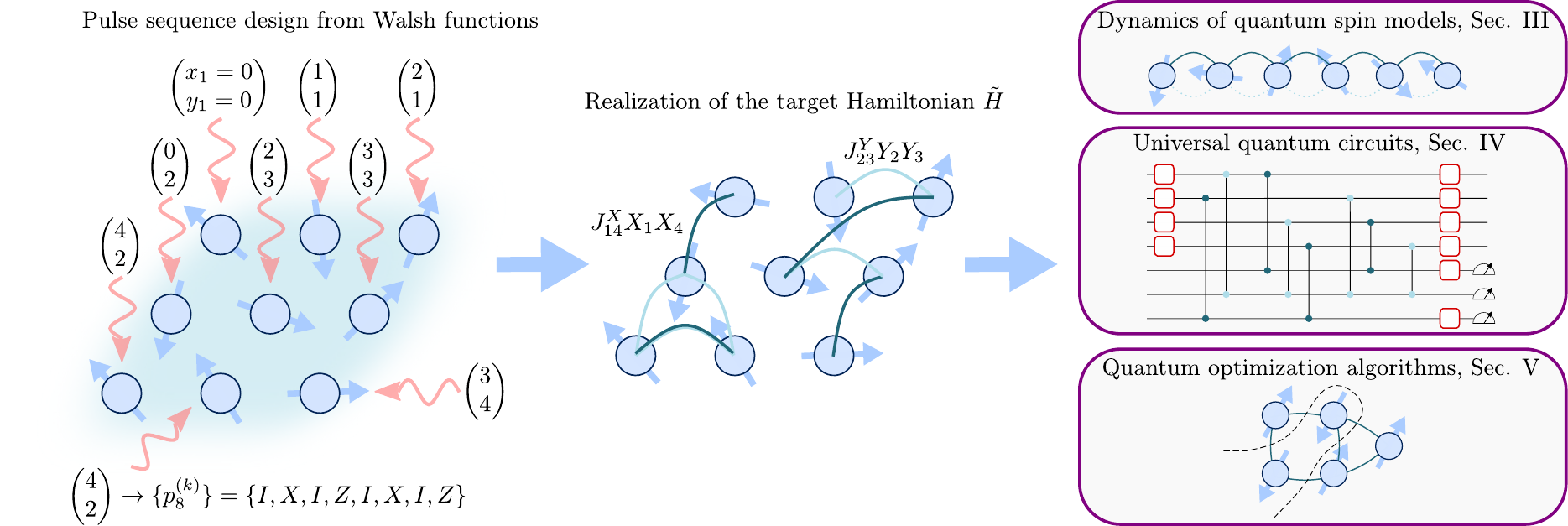}
    \caption{\textit{Universal quantum processors in spin systems via robust local pulse sequences} $-$ We illustrate the general features of the proposed protocol. 
    For each spin, two integer indices (i.e. the Walsh indices) $x_i$ and $y_i$ are chosen.
    Each couple of indices $(x_i, y_i)$ is mapped to a local pulse sequence using the Walsh functions language.
    The structure of the average Hamiltonian $\tilde{H}$ resulting from the pulse sequence is determined by the choice of indices $\{(x_i, y_i)\}$.
    I.e., whenever $x_i = x_j$ for two spins $i$ and $j$, $\tilde{H}$ will have a term $J^X_{ij}X_i X_j$, and whenever $y_i = y_j$ the Hamiltonian $\tilde{H}$ will have a term $J^Y_{ij}Y_i Y_j$.
    Alternating multiple pulse sequences of this kind can be used to engineer arbitrary Hamiltonians and quantum circuits.
    In the following sections, we demonstrate the robustness of this approach (Sec.~\ref{sec:robust}), and show applications to engineering dynamics of quantum spin models (Sec.~\ref{sec:ising_chain}), quantum error correcting protocols (Sec.~\ref{sec:surfacecode}) and quantum optimization algorithms (Sec.~\ref{sec:maxcut}). 
    }
\label{fig:summary}
\end{figure*}

Interacting atomic, molecular and solid-state spin systems are the fundamental constituents of various quantum processors. This includes in particular trapped ions~\cite{blatt2012quantum, monroe2021programmable}, Rydberg atoms~\cite{browaeys2020review, Morgado_2021_rydberg_review}, and superconducting circuits~\cite{devoret2013superconducting, gambetta2017building}.
In these systems, one typically combines the `native' interaction (resource) Hamiltonian $H_R$ with external control fields, in order to engineer a certain quantum dynamics described by a unitary operator $U$.
This is of key interest for building a quantum circuit in a quantum computer, where $U$ represents a sequence of quantum gates, but also for quantum simulation, where the aim is to engineer the dynamics $U=\exp(-i\tilde {H}T)$ of a many-body Hamiltonian $\tilde H$.
The goal of this article is to present a complete framework for quantum circuit and Hamiltonian engineering of generic quantum spin systems. Importantly, we show that this framework is at the same time \emph{flexible}, i.e. it can implement arbitrary pairwise interactions, and \emph{robust} with respect to various relevant experimental imperfections. 

In recent years, we have seen multiple proposals and experimental realizations of protocols where external driving fields engineer quantum dynamics from a spin system.
For instance, one can physically \emph{move} the spins to make them interact or decouple in a programmable fashion.
While this was demonstrated in remarkable pioneering experiments in Rydberg atom arrays~\cite{bluvstein_quantum_2022}, coherently moving the spins on short timescales, as well as exporting this method to other quantum platforms, are still significant technical challenges.
We will focus here on another approach that consists in applying external field `pulses' to the spins, in order to engineer an effective dynamics.

There are two main families of engineering protocols using pulses. 
First, in digital-analog quantum simulation/computing (DAQC)~\cite{hayes_programmable_2014, arrazola2016digital, lamata2018digital, Parra_Rodriguez_2020, gonzalez2021digital, garcia2023digital}, one aims to realize a large class of unitary operators $U$ given a single resource Hamiltonian $H_R$, see in particular Ref.~\cite{Parra_Rodriguez_2020, garcia2023digital} for numerical pulse sequence design to engineer universal quantum computation in spin systems. 
Note that, instead of addressing the spins individually, one can also use multiple spins to implement each qubit and drive them with a global field~\cite{Cesa_2023}.
While the DAQC approach is theoretically supported by theorems stating that we can always design a pulse sequence realizing a unitary operation $U$ to arbitrary precision using a fixed entangling Hamiltonian~\cite{dodd_universal_2002, wocjan2001universal,Wocjan_2002_theorems_networks}, 
strong experimental requirements are imposed.
In particular, one requires either to use pulses that are perfectly calibrated and quasi-instantaneous, as, in general, the interplay between errors arising from discretization of the dynamics and pulse calibration errors hinders a practical implementation~\cite{canelles2023benchmarking}, requiring either error mitigation techniques~\cite{garciamolina2022noise} or additional pulses~\cite{chevallier2024variational}.
The second family of protocols is related to the idea of Floquet engineering~\cite{Goldman_2014, Bukov_2015, Hung_2016_pcwg, choi2017dynamical, rajabi_dynamical_2019, Manovitz_2020_complex}.
Instead of aiming at universal quantum computation, here the effective dynamics is realized with a sequence of few pulses, designed mainly analytically and in a model-dependent fashion.
Importantly, when the pulse sequences are global, i.e. when all qubits are subject to the same pulses, one can make the resulting dynamics robust to experimental imperfections, such as external noise fields and errors due to imperfect pulses~\cite{choi_robust_2020, morong_engineering_2023, tyler2023higherorder, zhou2023robust}.
The simplicity and the robustness features have made this second approach appealing for quantum simulation experiments~\cite{Periwal_2021_cavity, Geier_2021_floquet_eng, scholl_microwave_2022, chen2023continuous, kranzl_experimental_2023, kranzl2023observation, nishad2023exchange_rydbergs}. 
 
In this work we demonstrate a protocol that merges the advantages of both DAQC and Floquet engineering, i.e. we provide a single experimental recipe to implement universal quantum processors based on arbitrary two-body interactions, using analytical pulse sequences that are inherently robust to experimental imperfections. 
This makes our proposal accessible in current setups, like Rydberg atom arrays and trapped ions. 
The use of Walsh functions~\cite{walsh1923closed, chien1975representations_walsh, beer1981walsh} is at the heart of our proposal.
This set of discrete orthogonal signal functions allows us to easily parametrize single qubit rotations both in space and time, and to obtain a simple Hamiltonian dynamics corresponding to the restriction of the resource Hamiltonian $H_R$ on a certain graph structure.
Based on simple graph decomposition arguments, we then show how to program multiple sequences of pulses to obtain an arbitrary two-body target Hamiltonian $\tilde H$. 
This allows in turn to realize arbitrary quantum computation, using for instance parallel two-qubit quantum gates. 
Similar pulse sequence design ideas based on Walsh functions have been previously considered to engineer Ising couplings~\cite{Leung00Hadamard, Bhole_2020, Tsunoda_2020, Rajakumar_2022}.
Importantly, we also show that the orthogonality property of Walsh functions offers `direct' analytical strategies for decoupling the system from unwanted sources of errors, without adding additional pulses. 
This is in contrast to previous DAQC protocols where these errors have been shown to be detrimental~\cite{canelles2023benchmarking}, and  makes our proposal promising for realizing quantum computation.
We analytically derive these robust pulse sequences in the case of imperfect pulses, as well as in presence of background fields. 

In Sec.~\ref{sec:pulse_sequence_design} we outline the general protocol and discuss how it can realize arbitrary quantum computation and simulation, along with a resource analysis in terms of number of pulses needed and errors, and finally present a controlled approximation scheme to reduce the number of pulses.
In Sec.~\ref{sec:robust} we translate robust strategies that were developed in the context of global pulses~\cite{choi_robust_2020} to the present case of locally-addressable pulses, and we derive analytically the robustness conditions for Walsh pulse sequences.
We then provide three examples showing the versatility of the protocol: engineering dynamics of quantum spin models beyond the resource Hamiltonian (Sec.~\ref{sec:ising_chain}), the implementation of a surface code (Sec.~\ref{sec:surfacecode}), and quantum optimization algorithms (Sec.~\ref{sec:maxcut}).
Finally, in Sec.~\ref{sec:experiments} we discuss in detail the possibility of implementing our protocol in state-of-art Rydberg atom arrays and trapped ions experiments.

\section{Introducing Walsh sequences}
\label{sec:pulse_sequence_design}

In this section we present our protocol in a way that is thought to be self-contained for most readers. 
We start by reviewing the pulse sequences approach to Hamiltonian engineering, in particular average Hamiltonian theory (AHT) and Trotter errors.
We then introduce the first key result of our work: pulse sequences parametrized with Walsh functions result in a simple structure of the engineered Hamiltonian.
This provides us with an experimental recipe, using multiple such pulse sequences, to engineer arbitrary target Hamiltonians (or quantum circuits).

\subsection{Pulse sequences for Hamiltonian engineering}
In this subsection we review the theory of Hamiltonian engineering based on pulses sequences. 
Let us consider for the moment a system consisting of $N$ spins $1/2$ (i.e., qubits) interacting via a static Hamiltonian $H_R$ (which we refer to as resource Hamiltonian), on which we can also apply pulses represented by instantaneous single-qubit gates $
    P = \bigotimes_{i=1}^N p_i$.
Note already that in an experimental setup, such gates can be approximately realized with external laser pulses, which are not instantaneous. 
We study the action of realistic pulses in Sec.~\ref{sec:robust}, and we derive conditions under which the instantaneous pulse approximation is accurate.
Commonly used pulses consist of Pauli rotations $p_i = R_{O}^{\alpha} = e^{-i\alpha O/2}$, $O \in \{X,\ Y,\ Z\}$ being a Pauli operator acting on the qubit $i$. For $\alpha = \pm \pi$, they amount to directly applying the operator $O_i$.
Several works about dynamical Hamiltonian engineering focus on the case in which $p_i = p_j \ \forall i,j$, which means that the pulse is acting in the same way on every qubit. We will refer to these pulses as global pulses.

In this work, we will focus instead on local $\pi-$pulses, where $p_i \in \{1, X, Y, Z\}$ and we can have $p_i \neq p_j$.
Let us now consider a pulse sequence $\{P^{(k)}\}=\{\bigotimes_{i=1}^N p_i^{(k)}\}$ of length $n$, which is repeated after a period of time $\tau$.
We partition the period $\tau$ in $n$ intervals of length $\tau/n$, and we apply the pulse $P^{(k)}$ at the beginning of the interval $k$, and the inverse pulse $(P^{(k)})^{-1}$ at the end of the interval $k$ (see Fig.~\ref{fig:pulses}).
The unitary evolution to which the system is subject over a period is
\begin{equation}
    \tilde{U}(\tau) = \prod_{k = 1}^n (P^{(k)})^{-1} e^{-i H_R \frac{\tau}{n}} P^{(k)}
\label{eq:Utau}
\end{equation}
where the product is applied right-to-left ($\prod_{k=1}^n A_k = A_n...A_2 A_1$).
Note that in an experiment, one may consider to merge the two pulses $(P^{(k-1)})^{-1}$ and $P^{(k)}$, and simply implement $(P^{(k-1)})^{-1}P^{(k)}$ at the time $\frac{(k-1)}{n}\tau$. However, it will be easier for us to derive robustness conditions in Sec.~\ref{sec:robust} keeping those two pulses as distinct objects.
Using the property of matrix exponentials $A^{-1} e^B A = e^{A^{-1}BA}$, we obtain
\begin{equation}
\label{eq:productformula}
    \tilde{U}(\tau) = \prod_{k = 1}^n e^{-iH^{(k)}\tau/n},\ H^{(k)} = (P^{(k)})^{-1} H_R P^{(k)}
\end{equation}
where $H^{(k)}$ are called the toggling-frame Hamiltonians.
The evolution of the system over a period is then described by a product formula.
Product formulas are connected to the evolution under a static Hamiltonian using the Floquet-Magnus expansion in small $\tau\to 0$~\cite{magnus1954exponential, kuwahara_floquetmagnus_2016, abanin2017effective}.
At the leading order, this expansion yields the average Hamiltonian theory (AHT), in which
\begin{equation}
    \tilde{U}(\tau) = e^{-i\tau\tilde{H}} + O(\tau^2),\ \tilde{H} = \frac{1}{n} \sum_{k=1}^n H^{(k)} 
\end{equation}
and $\tilde{H}$ is the average Hamiltonian.
To simulate a Hamiltonian dynamics up to a time $T$, usually the same product formula is applied for a fixed small time $\tau$, and then repeated $T/\tau$ times, $T/\tau$ being an integer.
The unitary error \mbox{$
    \mathcal{E}(\tau, T) := ||(\tilde{U}(\tau))^{T / \tau} -  e^{-iT\tilde{H}}||$}
can be quantified as~\cite{lloyd_universal_1996} 
\begin{equation}
\label{eq:lloyd_formula}
    \mathcal{E}(\tau, T) = \frac{\tau T}{2n^2} \Big{|} \Big{|} \sum_{k<l} [H^{(k)}, H^{(l)}] \Big{|} \Big{|} + O(\tau^2 T)
\end{equation}
where we take $|| \cdot ||$ to be the spectral norm, i.e. the largest singular value of the operator.
Eq.~\eqref{eq:productformula} is also referred to as the first-order $p = 1$ Trotter formula for $\tilde{H}$~\cite{childs_theory_2021}.
It is possible to engineer our pulse sequence to realize higher orders Trotter formulas, which feature a smaller error in simulating $\tilde{H}$.
For example, the $p=2$ Trotter formula requires a pulse sequence of length $n' = 2n$ (and the time duration of an interval is $\tau/n' = \tau/(2n)$) with $H^{(n + k)} \leftarrow H^{(n - k + 1)}$ for $k<n$, yielding $\mathcal{E} = O(\tau^2 T)$~\cite{Childs_2018_speedup, childs_theory_2021}.
$p>2$ Trotter formulas require to evolve the system with the negative resource Hamiltonian $-H_R$, which is not easily accessible in quantum simulation platforms.

\subsection{Hamiltonian engineering from Walsh functions}

In this subsection we start explaining the core ideas for our proposal.
Using the formalism of Walsh functions, we design pulse sequences $\{P^{(k)}\}$ that can realize target evolutions $\tilde H$.
From now on, we will consider a general $XY$ model as a resource Hamiltonian, i.e.
\begin{equation}
\label{eq:H_resource}
    H_R = \sum_{i<j} (J^X_{ij} X_i X_j + J^Y_{ij} Y_i Y_j) \\
\end{equation}
This covers in particular two very common types of interactions spin systems: Spin-exchange interactions $J^X_{ij}=J^Y_{ij}$ as for instance in dipolar Rydberg atom arrays, and Ising interaction $J^Y_{ij}=0$ as between trapped ions. 
More details about experimental implementation of such Hamiltonians combined with local pulses are given in Sec.~\ref{sec:experiments}.
While our analytical formulas will be general, we will consider for all examples power-law decaying spin-exchange interactions
\begin{equation}
\label{eq:LR_coupling}
    J^X_{ij} = J^Y_{ij} = - \frac{J}{r_{ij}^{\alpha}} \ .
\end{equation}
We now proceed by noting that, for our resource Hamiltonian Eq.~\eqref{eq:H_resource}, and using Pauli pulses $p_i^{(k)}\in\{1_i,X_i,Y_i,Z_i\}$ the expression of the toggling-frame Hamiltonians $H^{(k)}$ simplifies to
\begin{align}
    H^{(k)} &= \sum_{O = X, Y}\sum_{i<j}J_{ij}^{O}((p_i^{(k)})^{-1} O_i p_i^{(k)}) ((p_j^{(k)})^{-1} O_i p_j^{(k)})
    \nonumber \\
&=  \sum_{O = X, Y}\sum_{i<j}
J_{ij}^{O}
\left(
s_{i,O}^{(k)}
s_{j,O}^{(k)}
O_i
O_j
\right).
\end{align}
Here $s_{i,O}(k) = 1$ ($-1$) if $p_i$ commutes (anticommutes, respectively) with $O_i$.
Importantly, for each qubit $i$, it is easy to verify that we can always program the pulses $p_i^{(k)}$ to \emph{choose} the two signs $s_{i,X}^{(k)}$, $s_{i,Y}^{(k)}$
\begin{equation}
\begin{split}
\label{eq:walsh_rules}
p_i^{(k)}
&=
1_i\text{ if } s_{i,X}^{(k)}=s_{i,Y}^{(k)}=1
 \\
&=
X_i\text{ if } s_{i,X}^{(k)}=-s_{i,Y}^{(k)}=1
 \\
&=
Y_i\text{ if } -s_{i,X}^{(k)}=s_{i,Y}^{(k)}=1
 \\
&=
Z_i\text{ if } s_{i,X}^{(k)}=s_{i,Y}^{(k)}=-1.
\end{split}
\end{equation}

For a sequence of $k=1,\dots,n$ pulses, we can write the average Hamiltonian as 
\begin{equation}
\label{eq:Haverage}
    \tilde{H} = \sum_{i<j} ((s_{i,X}|s_{j,X})J^X_{ij}X_i X_j + (s_{i,Y}|s_{j,Y})J^Y_{ij}Y_i Y_j) \ .
\end{equation}
with the scalar product $(s_1|s_2)= \frac{1}{n}\sum_{k=1}^{n} s_1^{(k)}s_2^{(k)}$. 
The expression Eq.~\eqref{eq:Haverage} brings us to the key part of our proposal: As we can choose the signs $s_{i,X/Y}$ arbitrarily, let us define
$s_{i,X}^{(k)} = w_{x_i}^{(k)} \ ,\ s_{i,Y}^{(k)} = w_{y_i}^{(k)}$
where $w_a^{(k)}\in \{-1,1\}$, with $a$ integer and $k\in \{1,\dots,n\}$, are Walsh functions, which feature the orthonormality property $(w_a|w_b) = \delta_{a,b}$~\cite{walsh1923closed, chien1975representations_walsh, beer1981walsh} (see App.~\ref{app:walsh} for more details). 
In the case of Ising Hamiltonians, the same results hold by designing sequences of $X$ pulses using Walsh functions~\cite{Leung00Hadamard, Bhole_2020, Tsunoda_2020, Rajakumar_2022}.
We obtain the final form of $\tilde H$ which we will use throughout this work
\begin{equation}
\label{eq:single_walsh_h}
    \tilde{H} = \sum_{i<j} \left(G^X_{ij}J^X_{ij}X_iX_j + G^Y_{ij}J^Y_{ij}Y_iY_j\right)
\end{equation}
where $G^X_{ij}=\delta_{x_i, x_j}$ ($G^Y_{ij}=\delta_{y_i, y_j}$). 
This means that we can propose a general experimental recipe to program the pulses in such a way that two spins $i$ and $j$ interact via $X_iX_j$ couplings (or not) by assigning them the same Walsh index $x_i=x_j$ ($x_i\neq x_j$, respectively). The same result applies independently to $Y_iY_j$ couplings. This is illustrated in Fig.~\ref{fig:summary}. We show next the universality 
\begin{figure}[H]
    \centering
    \includegraphics[width=0.45\textwidth]{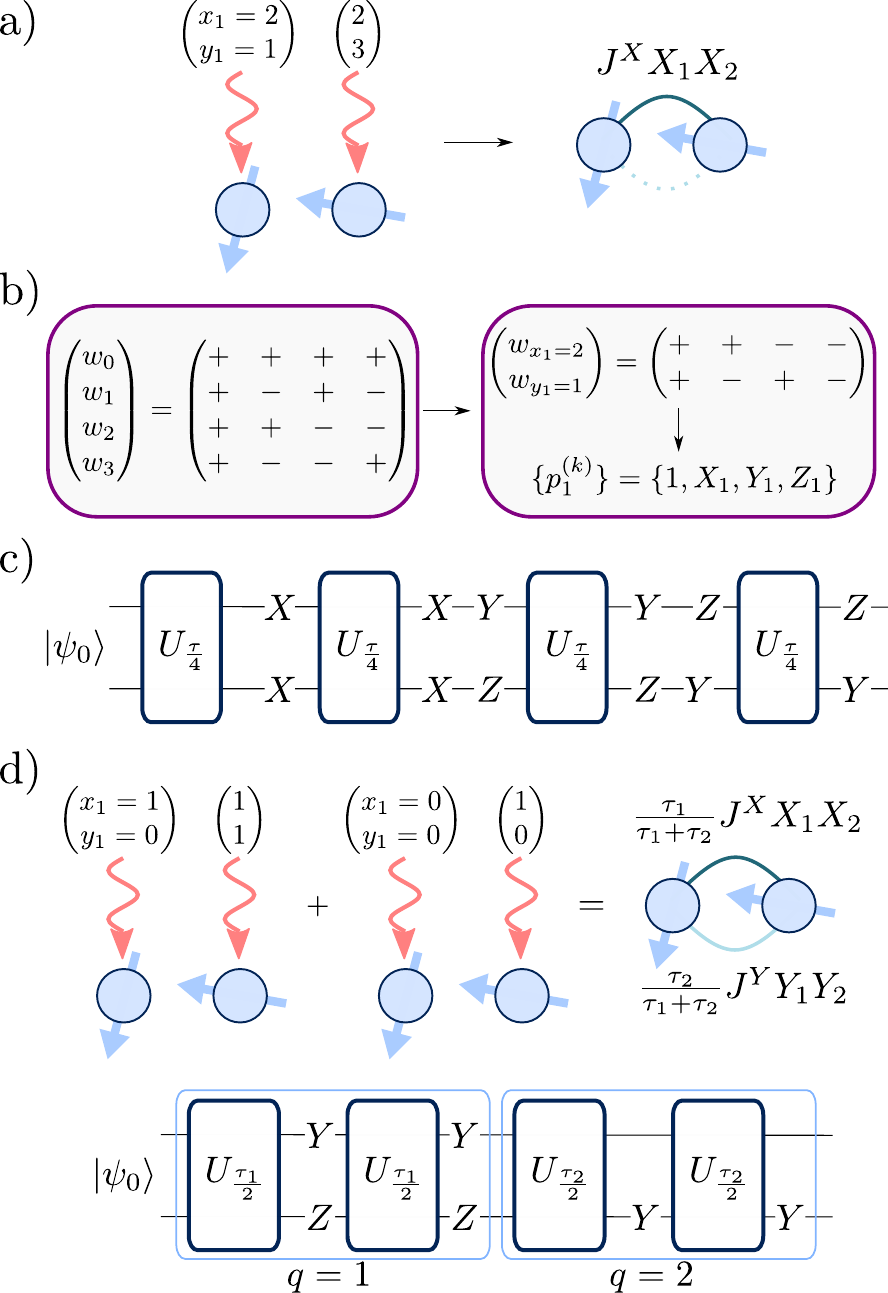}
    \caption{\textit{Pulse sequence design from Walsh functions} $-$ 
    a) We present a two-qubit example of pulse sequence design from Walsh functions. To each qubit, we associate two Walsh functions, one for the $XX$ interactions and the other for the $YY$ interactions. In particular, the choice here cancels the $YY$ interaction.
    b) The Walsh functions are obtained from a Hadamard matrix (App.~\ref{app:walsh}). Using the rules in Eq.~\eqref{eq:walsh_rules} we map the two Walsh functions for a given qubit in the resulting pulse sequence.
    c) The pulse sequence on the full system is given by the tensor product of the single-qubit pulse sequences.
    d) Multiple Walsh sequences can be alternated to engineer a more general Hamiltonian. In this example, we use $Q = 2$ Walsh sequences, the first one applied for a time $\tau_1$, and the second one applied for a time $\tau_2$.
    }
\label{fig:pulses}
\end{figure}
and flexibility of this approach (and later its robustness). 

It will be useful for what follows to use the language of graphs: $G^X_{ij}$ and $G^Y_{ij}$
are the adjacency matrix of the graph $\mathcal{G}_1^X $ ($\mathcal{G}_1^Y $).
We can view these two graphs as made with qubits as vertices, which are connected by a link if their Walsh indices match.
Formally, the links of these graphs are
\begin{equation}
\label{eq:graph_equation}
\begin{split}
    E(\mathcal{G}_1^X) = \{(i,j)\ |\ x_i=x_j\},\\ 
    E(\mathcal{G}_1^Y) = \{(i,j)\ |\ y_i=y_j\} \ .
\end{split}
\end{equation}
Note that we have obtained a mapping between a graph structure of interactions and a local pulse sequence which realizes it.
We will refer to each pulse sequence designed in this way as a Walsh sequence (or a Walsh pulse sequence, as in the title of the article). 
An explicit example for $N=2$ qubits is shown in Fig.~\ref{fig:pulses}~a)-c).
We choose $x_1 = x_2$ in order to couple the two qubits with respect to the $XX$ interaction, and $y_1 \neq y_2$ to decouple them with respect to the $YY$ interaction.
The choice of Walsh indices $(x_i, y_i)$ for each qubit associates two Walsh functions to it, from which we derive the resulting local pulse sequence $p_i^{(k)}$.
Finally, we note that the length $n$ of a Walsh sequence, i.e. the number of pulses to be applied in an experiment, is equivalent to the size of the Walsh function with the largest index, which is (see App.~\ref{app:walsh})
\begin{equation}
\label{eq:length_pulse_seq}
    n = 2^{\lceil \log_2 (\text{max}_i(x_i, y_i)+1)\rceil} \ .
\end{equation}
The longest Walsh sequence is obtained in the case all the qubits are required to be decoupled, for which $x_i=y_i=i$, which results in $n=O(N)$ (in particular, $N\le n < 2N$). 

\subsection{Universal quantum computing from Walsh sequences}
\label{sec:universal}

In this section we illustrate how to engineer a target spin Hamiltonian using $Q\ge 1$ Walsh sequences.
From the definition in Eq.~\eqref{eq:graph_equation}, we note that the interaction graphs $\mathcal{G}_1^X$ ($\mathcal{G}_1^Y$) cannot be arbitrary. For instance, one can readily see that it is not possible to couple two atoms $i_1$ and $i_2$ to a third atom $i_3$, without coupling $i_1$ to $i_2$.
In order to engineer more general average Hamiltonians, one thus needs to consider a pulse sequence made by alternating $q=1,\dots,Q$ Walsh sequences of length $n_q$, each one with a period $\tau_q$.
The global length of the resulting pulse sequence will be $n = \sum_{q=1}^Q n_q$, which thus scales at most as $n_q \le O(NQ)$.

In AHT, $Q$ Walsh sequences results in the unitary operator~\cite{childs_theory_2021, Childs_2018_speedup}
\begin{equation}
\label{eq:multiple_walsh}
    \tilde{U}(\tau) = \prod_{q=1}^Q \tilde{U}^{(q)}(\tau_q) = e^{-i\tau \sum_q \frac{\tau_q}{\tau} \tilde{H}^{(q)}} + O(\tau^{p+1})  \ .
\end{equation}
where the effective Hamiltonian $\sum_q (\tau_q/\tau) \tilde{H}^{(q)}$ can now be engineered for any possible spin Hamiltonian with two-body interactions, and thus in particular any quantum circuits of two-qubit gates (which is universal for quantum computing). 
While we report the complete proof of universality in App.~\ref{app:proof_universality}, let us sketch here the main idea. 
 
First, assuming here for the rest of this subsection $J_{ij}\neq 0$, and considering a target Hamiltonian of the form
\begin{equation}
    H_{\mathrm{T}} = \sum_{i<j}(\tilde{J}^X_{ij} X_i X_j + \tilde{J}^Y_{ij} Y_i Y_j  ),
\end{equation}
with $\tilde{J}^X_{ij},\tilde{J}^Y_{ij}\ge 0$, we define $g^O_{ij} := \tilde{J}^O_{ij} / J^O_{ij}$, which we decompose as 
\begin{equation}
\label{eq:decomposition}
g^O_{ij}=\sum_{q} \frac{\tau_q}{\tau} [G^O_q]_{ij}
\end{equation}
where $(G^O_q)_{ij}\in \{0,1\}$ are adjacency matrices of graphs whose structure is defined by Eq.~\eqref{eq:graph_equation}.
Therefore, we can implement each contribution $\propto [G^O_q]_{ij}$ with a Walsh sequence. 

The decomposition Eq.~\eqref{eq:decomposition} is always possible, and can be performed in various ways. The simplest situation arises when $g^{O}_{ij}$ is homogenenous ($g^{O}_{ij} \in \{0, 1\}$ $\forall i, j$).
This corresponds to
 various physically relevant scenarios both in quantum simulation and quantum computing, as we will illustrate below.
 In this case, we can simply decompose $g^{O}_{ij}$ in graphs of degree 1 with adjacency matrices $[G^O_q]_{ij}$ and $\tau_q=\tau$, i.e. in each such graph a spin $i$ is coupled to at most another spin.
Based on Eq.~\eqref{eq:single_walsh_h}, this corresponds to choosing for each $q$ a Walsh sequence $x_i=x_j$ ($y_i=y_j$) whenever $[G^X_q]_{ij}=1$ ($[G^Y_q]_{ij}=1$).
As shown in Sec.~\ref{sec:ising_chain},~\ref{sec:surfacecode}, this decomposition can be directly `seen' analytically in various relevant examples, when the number of Walsh sequences is $Q = O(1)$.
Alternatively, as shown in App.~\ref{app:proof_universality}, such decomposition can also be achieved in full generality by either
\begin{enumerate}
    \item starting from a decomposition of the complete graph~\cite{Bryant_2013, shyu2010decomposition, Galicia_2020_connectivity}, resulting in $Q=O(N)$ sequences.
    \item employing a numerical algorithm (see Algorithm~\ref{alg:graph_deco} in the appendix) to obtain a smaller number of Walsh sequences, i.e. $Q = O(d_\text{max})$, $d_\text{max}< N$ being the degree of the graph with adjacency matrix $g_{ij}^O$. 
\end{enumerate}

When $g^O_{ij}\notin \{0,1\}$ is not homogenous instead, the graph $[G^O_q]_{ij}$ needs to be simulated for different times $\tau_q$, with possible repeated links.
For instance, a strongly inhomogenenous and connected $g^O_{ij}$ can require to engineer as many Walsh sequences as the number of interacting links, resulting in the maximal value $Q=(N-1)N/2$.
This would imply that a random target Hamiltonian would require up to $\sum_{q=1}^Q n_q \lesssim O(N^3)$ pulses.
At the same time, a target Hamiltonian `commensurate' with the resource Hamiltonian, e.g. $g_{ij}^O$ with $O(1)$ different weights, will result in the much more favorable scaling of $\sum_{q=1}^Q n_q \lesssim O(Nd_{\text{max}})$ pulses.

Finally, in order to extend the construction above to more general couplings, it is sufficient to add a set of $Q$ pulses $\{P_{\text{set}}^q\}$, where $P_{\text{set}}^q$ is applied before the $q-$th Walsh sequence, and $\prod_{q=1}^QP_{\text{set}}^q = I$.
More details about this construction are given in App.~\ref{app:proof_universality}, and examples are given in the following sections.

To conclude, we briefly discuss the connection between Walsh pulse sequences and quantum circuits.
In fact, any layer of a quantum circuit made of two-qubit gates acting on different pairs of qubits can be realized by the dynamics of a single pair-wise interacting Hamiltonian $\tilde{H}$.
Since the connectivity of such a Hamiltonian will be represented by a graph of degree 1, this set of gates can be implemented \emph{in parallel} via Walsh sequences (or a single Walsh sequence, in some cases).
Each one of this Walsh pulse sequences will require $n = O(N)<2N$ pulses, and they will be applied in total $T/\tau$ times, where $\tau$ fixes the Trotter error, and $T$ is the longest evolution time needed by a two-qubit gate to be realized.
An explicit example will be discussed in Sec.~\ref{sec:surfacecode}.
Finally, let us note that allowing for groups of $N_g>2$ qubits to interact might be used to realize $N_g-$qubit gates.
This implies that not only Walsh pulse sequences can find application in realizing quantum gates in systems with collective interactions, but also that these interactions could be exploited to realize novel gates, using our framework.

\subsection{Trotter errors}
We now assess the Trotter errors from realizing the average Hamiltonian $\tilde{H}$ with $Q$ Walsh sequences, each one with a period $\tau_q$.
Starting from Eq.~\eqref{eq:lloyd_formula}, we derive a bound for unitary Trotter errors for one-dimensional systems with power-law decaying interactions as in Eq.~\eqref{eq:LR_coupling}, which reads for $p=1$ (see App.~\ref{app:errors})
\begin{equation}
\label{eq:trotter_bound}
    \begin{cases}
    \mathcal{E}(\{\tau_q\}, T) \leq a_{\alpha} N(JT)(J\sum_q\tau_q) \ \text{if} \ \alpha > 1 \\
    \mathcal{E}(\{\tau_q\}, T) \leq b_{\alpha} N^{3 - 2\alpha}(JT)(J\sum_q \tau_q) \ \text{if} \ \alpha < 1
\end{cases}
\end{equation}
where $a_{\alpha}, b_{\alpha} = O(1)$ are functions of only $\alpha$. 
Therefore, to simulate a dynamics with a Trotter error $\mathcal{E} \leq \epsilon$, we can fix the period of the total cycle to be $J\sum_q\tau_q = \kappa_{\epsilon}N^{-1}(JT)^{-1}$ for $\alpha > 1$, or $J\sum_q\tau_q = \kappa_{\epsilon}N^{2\alpha - 3}(JT)^{-1}$ for $\alpha < 1$, $\kappa_{\epsilon} = O(\epsilon)$ being a constant that can be tuned to obtain the desired error value.
We believe this can be generalized easily to higher spatial dimensions and Trotter orders $p>1$, using the techniques presented in Ref.~\cite{childs_theory_2021}.
This estimate serves as a performance guarantee for a generic Hamiltonian engineering protocol, and shows that the required number of number of pulses is polynomial with system size.

\subsection{Reducing the number of pulses with a cut-off distance}
\label{subsec:cutoff}

We conclude this section by presenting a controlled approximation scheme which allows to reduce the number of pulses $n$ in Walsh sequences.
This is important in light of an experimental implementation since upon fixing the physical limit for pulse repetition frequency $\tau/n$, decreasing $n$ allows us to reduce $\tau$, and therefore Trotter errors.
In particular, let us consider the situation in which we have a target Hamiltonian with finite-range interactions, i.e. it exists $r = O(1)$ such that $\tilde{J}^O_{ij} = 0$ for $r_{ij}>r$ for $O = X, Y$.

Let us consider the associated Walsh sequence with length $n = O(N)$.
The approximation which we rely on consists in choosing a cut-off distance $\Lambda_r$ for which we assume $J^O_{ij} \approx 0$ for $r_{ij}>\Lambda_r$ in the resource Hamiltonian.
For one-dimensional systems, it is possible to prove that, under this assumption, the Walsh indices can be chosen to be bounded as $x_i,y_i< \Lambda_w$, with 
\begin{equation}
    \Lambda_w := \text{max}_i (x_i, y_i) < 2\Lambda_r,
\end{equation}
which corresponds to the largest index of the `compressed' Walsh sequence, of which the length is given by Eq.~\eqref{eq:length_pulse_seq}. 
A proof of this statement and a Walsh indices assignment algorithm are given in App.~\ref{app:errors2}.
We believe that a similar approach can also be adapted to higher dimensions.

Let us now quantify the errors made by the cutoff approximation of the resource Hamiltonian. We have 
\begin{equation}
    (\tilde{U}(\tau))^{T/\tau} = e^{-iT(\tilde{H} + H_E)} + O(\tau^p T)
\end{equation}
where
\begin{equation}
    H_E = \sum_{i, j, |i-j|>\Lambda_r}(J^X_{ij}\delta_{x_i,x_j}X_i X_j + J^Y_{ij}\delta_{y_i,y_j} Y_i Y_j) 
\end{equation}
is a perturbation to our target Hamiltonian $\tilde H$.
For small enough $\tau||H_E||$, the unitary error induced by $H_E$ in a single Trotter step is
\begin{equation}
    \tilde{U}(\tau) - e^{-i\tau\tilde{H}} =   \tau H_E e^{-i\tau\tilde{H}} + O(\tau^2)
\end{equation}
and therefore if $|| H_E || = O(\tau)$ the residual error will be still proportional to $p=1$ Trotter errors.
For $J^O_{ij} = J/|i-j|^{\alpha}$, $\alpha>3$, we provide an analytical guarantee that the errors given by the choice of a cut-off can be made arbitrarily small.
In particular, we find that $||H_E|| = \epsilon \tau$ is ensured by choosing $\Lambda_w = O((N/\epsilon)^{\frac{2}{\alpha - 1}})$.
We provide a numerical example for the cut-off scheme in Sec.~\ref{sec:ising_chain} for $\alpha = 0.2,\ 3$, finding that for $\alpha = 3$ the cut-off errors can be made much smaller that Trotter errors.

\section{Walsh sequences are robust against experimental imperfections}
\label{sec:robust}
The formalism of Walsh sequences presented in the previous section provides us with analytical pulse sequences to decouple/couple selectively groups of qubits to engineer specific Hamiltonian terms.
Let us show now that the use of Walsh functions, based on similar `decoupling strategies', also allows us to suppress the effect of experimental imperfections.
This step is essential to implement reliable quantum computation and simulation based on pulse sequences, as alternating global Hamiltonian evolution for short times with single-qubit imperfect pulses can yield a quick build-up of errors~\cite{canelles2023benchmarking}.
In the following, we provide analytical evidence of robustness of our protocol with respect to rotation angle errors, finite pulse duration, and presence of background magnetic fields.
In particular, the latter situation will be useful to draw connections between our formalism and the research topic of dynamical decoupling. 

\subsection{Rotation angle errors: From single averaging to double averaging}
\label{subsec:rotation_errors}

Let us first consider errors in the rotation angle applied by the pulses during one Walsh sequence.
This kind of errors can be related to a miscalibration of the single-qubit pulses~\cite{canelles2023benchmarking}, or to over- or under-illumination of certain qubits by the applied laser~\cite{kranzl_experimental_2023, kranzl2023observation}.
We model a pulse with rotation angle errors as
\begin{align}
    p_i^{(k)} = e^{-is_i(\pi + \delta_i)[O]_i/2},
\end{align}
where $s_i = \pm 1$ when the pulses are implemented as a $\pi$ rotation ($-\pi$ rotation, respectively), $\delta_i$ is a single-qubit rotation angle error, and $[O]_i \in \{1_i, X_i, Y_i, Z_i\}$ represents the pulse we wish to apply on the $i-$th qubit (i.e., $p_i^{(k)}=[O]_i$ in absence of errors $\delta_i=0$). In the following, we will take advantage of the freedom to choose the type of pulses $s_i$ to design robust Walsh sequences.

As shown in App.~\ref{app:rotation_angle_errors}, the effect of rotation errors 
under Walsh sequences leads at order $O(\delta)$ to a remarkably simple expression of the average Hamiltonian $\tilde H+\tilde{H}_{\text{RA,err}}$, with 
\begin{align}
    \tilde{H}_{\text{RA,err}} &= \sum_{O = X, Y}&\sum_{i<j}J^O_{ij}\left(\delta_i s_i \mathcal{O}_i^j O_j + \delta_j s_j O_i \mathcal{O}_j^i \right),
    \label{eq:HRAerr}
\end{align}
and
\begin{equation}
\begin{split}
    \mathcal{X}_i^j =& \frac{1}{n}\sum_{k=1}^n \frac{1 - w_{x_i}^{(k)}}{2}w_{x_j}^{(k)}\left(\frac{1 - w_{y_i}^{(k)}}{2} Y_i - \frac{1 + w_{y_i}^{(k)}}{2}Z_i \right) \\
    \mathcal{Y}_i^j =& \frac{1}{n}\sum_{k=1}^n \frac{1 - w_{y_i}^{(k)}}{2}w_{y_j}^{(k)} \left(\frac{1 + w_{x_i}^{(k)}}{2} Z_i - \frac{1 - w_{x_i}^{(k)}}{2}X_i \right) \ .
\end{split}
\end{equation}

The simple form of Eq.~\eqref{eq:HRAerr} suggests to build a robust sequence using a second averaging strategy~\cite{choi_robust_2020, haeberlen1971resonance} based on varying with time the sign functions $s_i$. 
Let us first recall that in the previous section, we considered repeating the same Walsh sequence identically $T/\tau$ multiple times to propagate the Hamiltonian until a certain evolution time $T$. To make this evolution robust, we will realize instead the evolution 
\begin{equation}
\underbrace{
    (\tilde{U}(\tau))^{T/\tau}
    }
    _{\text{single averaging}}
    \rightarrow
    \underbrace{\prod_{l=1}^{T/\tau}\tilde{U}^{(l)}(\tau)}
    _{\text{double averaging}}.
\end{equation}
where the sign $s_i^{(l)}$ becomes a function of the index $l$, and we define them to be Walsh functions as $s_i^{(l)} = w_{e_i}^{(l)}$. Here we choose
$e_i>0$. Note that this second averaging does not require the introduction of additional pulses in our protocol.
In this case, averaging the expression Eq.~\eqref{eq:HRAerr} of the error $\tilde{H}_{\text{RA,err}}$ over the $T/\tau$ evolutions, one obtains that the average error Hamiltonian  at order $O(\delta)$ \emph{vanishes exactly}, i.e. $\sum_{l=1}^L\tilde{H}_{\text{RA,err}}^{(l)} = 0$, $L$ being the periodicity of the sign function with the largest Walsh index $e_i$.
This demonstrates how a simple choice of pulsing angles can make Walsh pulse sequences resilient to calibration errors.
We report in App.~\ref{app:rotation_angle_errors} numerical data illustrating the use of robust Walsh sequences in the presence of rotation angle errors.

\subsection{Double averaging finite pulse duration errors}
\label{subsec:finite_pulses}

So far we have considered the qubits to be not evolving with the resource Hamiltonian during the application of the pulses. 
While this assumption is certainly valid in ion chains, where laser mediated interactions can be switched on and off, the situation may be different in Rydberg atom arrays in the `frozen gas' regime, where interactions are always on, see Sec.~\ref{sec:experiments}.
At the same time, it might be desirable to keep the evolution under the resource Hamiltonian on even during the application of pulses, to avoid ramp-up and ramp-down errors~\cite{Parra_Rodriguez_2020, canelles2023benchmarking}.
Let us consider thus the effect of the finite duration of a pulse and define the single-qubit Hamiltonian term 
\begin{equation}
    H_p = \frac{\pi}{2t_p}\sum_i s_i [O]_i
\end{equation}
generating the pulse $P^{(k)} = e^{-it_pH_p}$, where $t_p$ is the time duration of a pulse, and we assume to a have a square-wave pulse.
So, during the interval $k$, the system is subjected to the following unitary evolution
\begin{equation}
    U^{(k)}(\tau) = e^{-it_p(H_R - H_p)}e^{-i\left(\frac{\tau}{n} - 2t_p \right)H_R}e^{-it_p(H_R + H_p)} \ .
\end{equation}
In the limit of an infinitely short pulse $t_p=0$, we recover the idealized limit of the previous section, c.f. Eq.~\eqref{eq:Utau}.
Upon defining the finite pulse duration error parameter 
    $\epsilon_{\text{FP}}=2nt_p/\tau$, 
which amounts to the time portion of the interval $k$ during which pulses are applied, we obtain the following average Hamiltonian (see App.~\ref{app:finite_pulse_duration_errors} and Refs.~\cite{Haeberlen68, choi_robust_2020})
\begin{align}
    \tilde{H}_{\text{FP}} &= \left(1 - \epsilon_{\text{FP}} \right)\tilde{H}\ +  \nonumber \\
    &+\epsilon_{\text{FP}}\frac{1}{n}\sum_{k=1}^n 
    \frac{1}{t_p}\int_0^{t_p}dt\ (Q^{(k)}(t))^{-1} H_R Q^{(k)}(t)
\end{align}
where $Q^{(k)}(t) = e^{-itH_p}$ can be interpreted as a partially applied pulse.
Notice that we can repeat the above treatment for the case of pulses with different wave-shapes.

As for rotation errors, the expression of the average Hamiltonian takes a simple form when parametrizing the pulses with Walsh functions (under the assumption here that $x_i\neq 0$, $y_i\neq 0$), see App.~\ref{app:finite_pulse_duration_errors}. 
When using a second averaging with time changing signs $s_i^{(l)}=w_{e_i}^{(l)}$, and $e_i\neq e_j\neq 0$ we first obtain
\begin{equation}
\begin{split}
    \tilde{H}'_{\text{FP}} =\left(1 - \frac{5 \epsilon_{\text{FP}}}{8} \right)\tilde{H} + \frac{3\epsilon_{\text{FP}}}{8}H_R \ .
\end{split}
\end{equation}
At this point, it is possible to further deform the Walsh sequence to simulate $\tilde{H}$ in AHT.
This is achieved by 1) realizing the interval $k=1$, the ones in which $H^{(k=1)} = H_R$, for a reduced time $\frac{\tau}{n} - \frac{3nt_p}{4}$, and 2) correcting the total simulation time $T$ as $T \rightarrow T\left(1 - \frac{5 \epsilon_{\text{FP}}}{8}\right)^{-1}$.
These robustness conditions, and time deformations, completely remove any contribution in AHT of the finite pulse duration effects, and at the same time they remove the dominant contribution in AHT for rotation angle errors.
To corroborate our analysis, a numerical example is given in Sec.~\ref{sec:ising_chain} and App.~\ref{app:rotation_angle_errors}, showing both the correction of finite-pulse duration effect alone, and together with rotation angle effects.

\subsection{Suppression of background magnetic fields and connections to dynamical decoupling}
\label{subsec:dyndec}

As a last source of errors, we consider single-qubit noise in the resource Hamiltonian.
To decouple a qubit system from single qubit noise (or other sources of unwanted interaction) is referred to in the literature as dynamical decoupling~\cite{Vandersypen_2005_nmr_review, suter_2016_dyndec_review}.
Recently, dynamical decoupling has been connected to Hamiltonian engineering~\cite{morong_engineering_2023, kranzl_experimental_2023, kranzl2023observation}, in order to increase accuracy in quantum simulation protocols beyond the simulation of the resource Hamiltonian. Here, we show that this idea further extends to the case of universal quantum evolutions treated here. 

We consider a resource Hamiltonian which also has space-dependent unwanted single-qubit fields
\begin{equation}
    H_R' = H_R + H_{\text{ext}}
\end{equation}
with
\begin{equation}
    H_{\text{ext}} = \sum_i (h^x_i X_i + h^y_i Y_i + h^z_i Z_i)
\end{equation}
which can model disorder fields, or slowly-varying noise.
After a Walsh sequence, we obtain again a simple expression (see App.~\ref{app:dynamical_decoupling})
\begin{equation}
    \tilde{H}_{\text{ext}} = \sum_i ( \delta_{0, x_i} h^x_i X_i + \delta_{0, y_i} h^y_i Y_i + \delta_{x_i, y_i} h^z_i Z_i) \ .
\end{equation}
In this case, there is no need for double averaging: Full decoupling can be then achieved after a single Walsh sequence by 1) not using the Walsh function $w_0^{(k)}$ and 2) using different Walsh indices for the $X$ and $Y$ operators on the same qubit.
This can be always be achieved by enumerating the $y_i$ starting from the number after the largest $x_i$, which doubles the number of Walsh indices (hence the length of the resulting Walsh sequences). Note that this robustness condition can be applied simultaneously with the ones described in the two previous subsections. 

\section{Application to Hamiltonian engineering}
\label{sec:ising_chain}

\begin{figure*}[t]
    \includegraphics[width=0.99\textwidth]{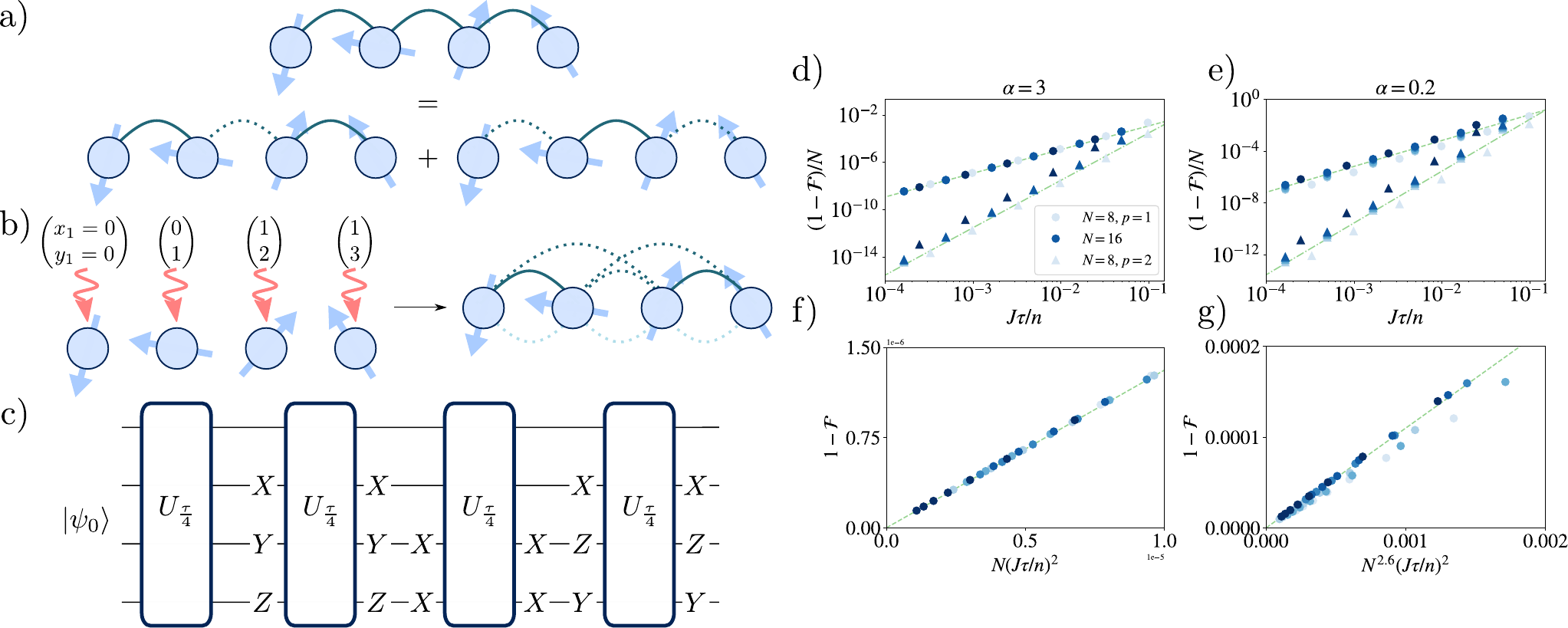}
    \caption{\textit{Ising dynamics from Walsh sequences} $-$
    a) We show how the interaction rescaling $g^X_{ij}$ is realized with $Q = 2$ Walsh sequences from the decomposition $g^X_{ij} = [G^X_1]_{ij} + [G^X_2]_{ij}$.
    b) We show a possible assignment of Walsh indices for $[G^X_1]_{ij}$, and c) the resulting Walsh sequence, where $U_{\frac{\tau}{n}} = e^{-iH_R\tau/n}$. 
    d)-g) We plot the fidelity error $1 - \mathcal{F}$ between the output state $\ket{\psi_{\text{sim}}}$ for a given timestep $\tau/n$, and the ideal state $\ket{\psi_{\text{cluster}}}$ for $T = \pi/(4J)$, $T/\tau$ being an integer and $n$ the length of the Walsh sequences. 
    d),e) We show the fidelity error per qubit as a function of the timestep $\tau/n$ and the number of qubits $N$. The round (triangular) markers refer to the dynamics realized by the $p = 1$ ($p = 2$) Trotter formula, and the dashed (dash-dotted) lines being guides to the eye for $\sim (J\tau/n)^2$ ($\sim (J\tau/n)^4$). 
    f),g) We note a data collapse of the fidelity error with $N(J\tau/n)^2$ ($N^{3 - 2\alpha}(J\tau/n)^2$) for $\alpha>1$ ($\alpha<1$). 
    In d),f) we choose $\alpha = 3$ as in dipolar Rydberg atom arrays, in e),g) $\alpha = 0.2$ as realizable in trapped ion chains. } 
\label{fig:clusterstates}
\end{figure*}

We are now ready to illustrate examples of our approach. In what follows, we consider a long-range resource Hamiltonian, with couplings defined from Eq.~\eqref{eq:LR_coupling}.
We start with the implementation of the Ising model with nearest-neighbor couplings, used as a paradigmatic example to benchmark the performances of our approach. This is for instance related to the problem of reducing interactions from long-range to short-range in quantum simulators~\cite{Lee_2016}.
This is also useful in quantum computing, as the model allows us to prepare graph states, in particular cluster states~\cite{briegel_persistent_2001, verresen2022efficiently}.

Our target model is defined 
as
\begin{equation}
\label{eq:NN_Ising}
    H_{\text{Ising}} = -J\sum_{i=1}^{N-1} X_i X_{i+1}.
\end{equation}
This corresponds to the two following rescaled interactions defined in Sec.~\ref{sec:universal}
\begin{equation}
    g_{ij}^X = \delta_{i, j\pm 1},\ g_{i j}^Y = 0 \ .
\end{equation}
which represent a homogenenous graph of degree 2, and can be straightforwardly decomposed in two graphs of degree $1$
\begin{equation}
    [G_{q = 1}^X]_{2i-1, 2i} = 1,\ [G_{2}^X]_{2i, 2i+1} = 1
\end{equation}
while all the other terms are zero.
Finally, $[G_{1}^X]_{i, j}$ can be realized by choosing the Walsh indices $x_i = \lfloor (i-1)/2 \rfloor$, and $[G_2^X]_{ij}$ by choosing $x_i = \lfloor i/2 \rfloor$.
$[G_{q}^Y]_{i, j}$ instead are always realized with $y_i = i-1$.
In Fig.~\ref{fig:clusterstates}a-c) we show an example of the decomposition and Walsh indices assignment for $N = 4$.

We benchmark numerically the protocol in Fig.~\ref{fig:clusterstates}d)-g).
In particular, we assess Trotter errors numerically studying the state fidelity $\mathcal{F}=|\langle\psi_{\text{cluster}}|\psi_{\text{sim}}\rangle|$ for $T = \pi / (4J)$ and initial state $\ket{\psi_0} = \bigotimes_i \ket{0}_i$.
This choice outputs cluster states $\ket{\psi_{\text{cluster}}}$ in the case of perfect Ising dynamics ($\tau = 0$)~\cite{briegel_persistent_2001, verresen2022efficiently}.
The output state fidelity is related to the unitary Trotter errors bounded by Eq.~\eqref{eq:trotter_bound}, since the latter is a bound for the trace distance between the simulated and the target states~\cite{hayes_programmable_2014}, which for pure states coincides with the following fidelity error~\cite{wilde2019shannon}
\begin{equation}
    ||(\tilde{U}(\tau))^{T/\tau} - e^{-iT\tilde{H}}|| \geq \sqrt{1 - \mathcal{F}} \ .
\label{eq:fide_trotter}
\end{equation}
To check the output state fidelity, we numerically simulate the pulse sequence protocol using exact numerical methods, with size varying from $N=8$ to $N=16$, and various periods $\tau$.
We consider the two cases $\alpha = 3$ and $\alpha = 0.2$.
The results are shown in Fig.~\ref{fig:clusterstates}d)-g).
In panels d,e), we plot the fidelity error per qubit $(1 - \mathcal{F})/N$, for pulse sequences that correspond to $p = 1$ (round markers) and $p = 2$ (triangular markers) Trotter formulas.
For the behavior with the timestep $\tau$, we observe a power-law decay of the fidelity error per qubit $\sim \tau^{2p}$, which is for $p=1$ consistent with the upper bounds of  Eqs.~\eqref{eq:trotter_bound}-\eqref{eq:fide_trotter}.
For the behavior with the size $N$ instead, in the case $\alpha = 3$ and $p = 1$ the fidelity error per qubit appears to be independent of the size $N$, whereas for $\alpha=0.2$ it increases with $N$.
To better understand the scaling with $N$, we consider the case $p = 1$, where can expect scalings of at most $\sim N^2$ for $\alpha=3$ and $\sim N^{2(3 - 2\alpha)}=N^{5.2}$ for $\alpha =0.2$, based on the results of Eq.~\eqref{eq:fide_trotter} and Eq.~\eqref{eq:trotter_bound}.
By performing a data collapse, we observe a scaling of the fidelity error of $\sim N$ for $\alpha =3$ and $\sim N^{2.6}$ for $\alpha=0.2$.
\begin{figure}[H]
    \centering
    \includegraphics[width=0.47\textwidth]{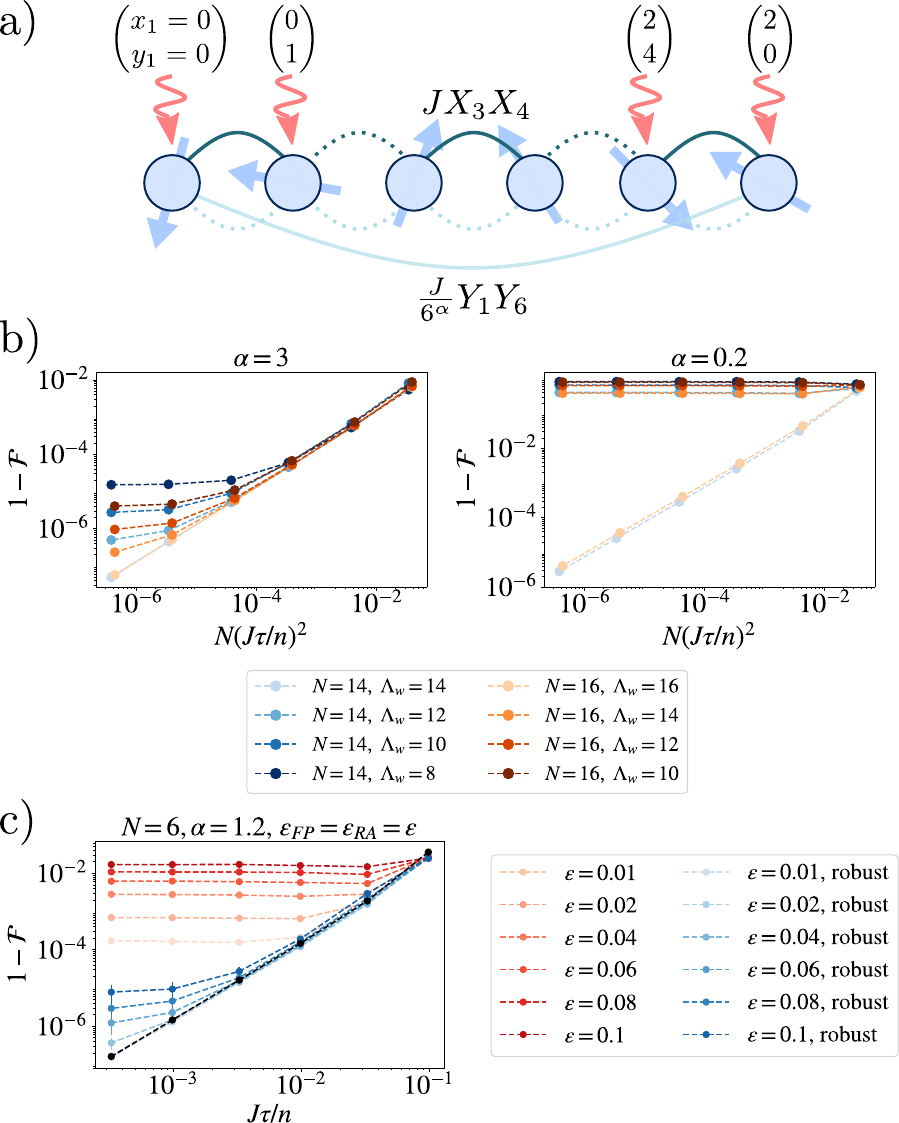}
    \caption{\textit{Ising dynamics with a cut-off approximation and robustness conditions} $-$ a) A Walsh sequence to engineer the Ising chain on $N=6$ qubits, if imposed a cut-off on the largest Walsh index $\Lambda_w=4$, results in a further unwanted coupling $\frac{J}{6^{\alpha}}Y_1 Y_6$, with resource interactions described by Eq.~\eqref{eq:LR_coupling}. 
    b) We numerically check the effect of a cut-off $\Lambda_w$ on the Ising dynamics at time $T = \pi / (4J)$. We plot the fidelity error for $\alpha = 0.2, 3$, and chains of $N = 14, 16$ qubits. For $\alpha = 3$ we observe the fidelity error decreasing as Trotter errors until a given value of $\tau$, for which it saturates. For $\alpha = 0.2$ the saturation happens for any $\tau$ if $\Lambda_w < N-1$.  
    c) We study the effects of pulse imperfections in the same engineered Ising dynamics for a chain of $N = 6$ qubits, with $\alpha = 1.2$. 
    We consider systematic rotation angle error $\delta_i$ on the $i-$th qubit, $\delta_i$ sampled uniformly in $[-2\epsilon_{\text{RA}}J, 2\epsilon_{\text{RA}}J]$ (datas are averaged over 64 samples) and pulses with duration $t_p = \frac{\tau}{2n}\epsilon_{\text{FP}}$.
    The red lines represent the fidelity error for bare pulse errors, while for the blue lines robustness conditions of Sec.~\ref{sec:robust} are met. The black lines represent only Trotter errors, i.e. absence of pulse imperfections.
    We observe how robustness conditions reduce pulse-related fidelity errors by orders of magnitude.
    }
    \label{fig:cutoff}
\end{figure}
This shows that the Trotter bound predicted is loose in this case.

In the case presented, the length of the pulse sequence considered was $n=N$, which was required to cancel completely the $YY$ component of the resource Hamiltonian. As seen in Sec.~\ref{subsec:cutoff}, we can reduce the pulse sequence length drastically using a cut-off distance $\Lambda_r$ above which we can neglect interactions. 
Using the Walsh indices $x'_i = x_i\ \text{mod}\ \Lambda_r$ ($y'_i = y_i\ \text{mod}\ \Lambda_r$), we obtain the fidelity shown in Fig.~\ref{fig:cutoff}b).
As expected, for the short-range case $\alpha = 3$, the introduction of a cut-off $\Lambda_r<N$ induces a very modest error $1 - \mathcal{F} \sim 10^{-5}$ for the worst case considered $N = 14$, $\Lambda_w = 8$. 
For the long-range case $\alpha = 0.2$, the notion of the cut-off distance is meaningless, since, as soon as $\Lambda_r<N-1$, the fidelity error becomes independent of $\tau$ and of order one.

Finally, we check numerically the effect of pulse induced errors, and the performances of the robustness conditions presented in Sec.~\ref{sec:robust}.
We plot the results in Fig.~\ref{fig:cutoff}c).
While the black lines represent the fidelity error for perfect pulses, the red and blue lines represent the fidelity error when pulse errors are present (see Apps.~\ref{app:rotation_angle_errors}-\ref{app:finite_pulse_duration_errors} for the effects studied separately).
We model the rotation angle errors by sampling a qubit-dependent rotation angle error of $\delta_i$, which we sample from a uniform distribution in the interval $[-2\epsilon_{RA} J, 2\epsilon_{RA} J]$.
At the same time, we consider a finite duration for the pulses of $t_p = \frac{\tau}{2n}\epsilon_{\text{FP}}$.
The fidelity errors are plot for various values of the errors $\epsilon_{\text{RA}} = \epsilon_{\text{FP}} = \epsilon$, $N = 6$ qubits and $\alpha = 1.2$.
We observe that the introduction of the robustness conditions using the double averaging strategy discussed in Sec.~\ref{sec:robust} (blue lines) allows to reduce the fidelity error of several orders of magnitude compared to a single averaging (red lines). 

Note that this example for the Ising model can be straightforwardly extended to the situation of translation-invariant $XXZ$ type Hamiltonians on arbitrary spatial geometry with finite connectivity. This can for example be done by implementing separately the $XX$, $YY$ and $ZZ$ interactions on degree 1 graphs. This idea can be applied in particular to build Floquet Kitaev models, which are relevant for topological quantum computing with non-abelian anyons~\cite{Sun_2023_kitaev_floquet, Kalinowski_2023_kitaev_floquet}.

\section{Application to parallel quantum circuits and quantum error correction}
\label{sec:surfacecode}

\begin{figure*}[t]
    \includegraphics[width=0.99\textwidth]{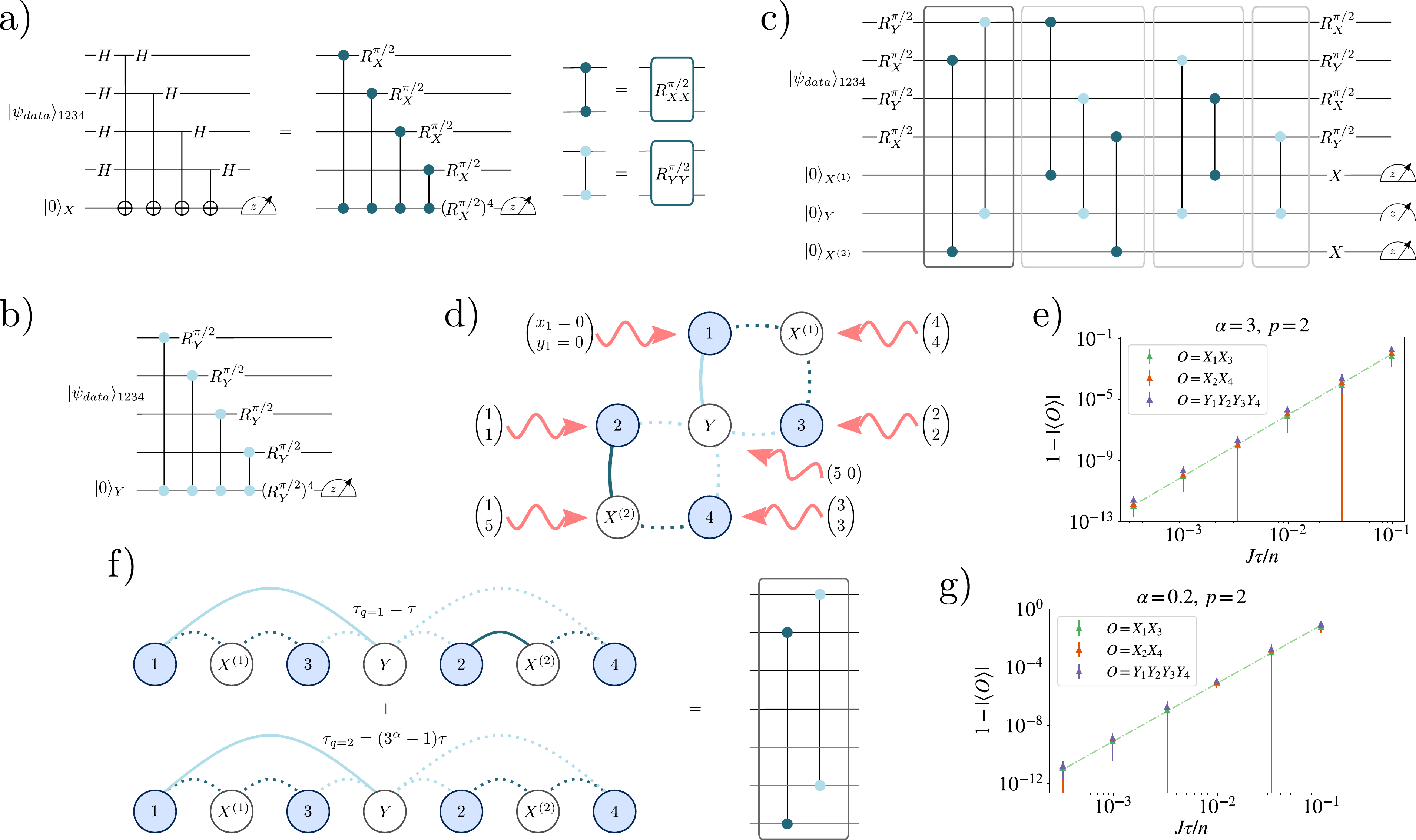}
    \caption{\textit{Implementing the 7-surface code with Walsh sequences} $-$ a,b) The building block of the surface code circuit, originally made by Hadamard and CNOT gates, can be mapped to a circuit made of two- and single-qubit rotations.
    The two-qubit rotations can be realized using $XX$ and $YY$ Ising interactions.
    c) This makes it possible to map the whole surface code quantum circuit to 4 layers of parallel $XX$ and $YY$ Ising interactions, with Ising interactions run for a time $T = \pi/(4J)$.
    Each one of these blocks can be realized by repeating a single Walsh sequence, e.g. in d) we illustrate the choice of Walsh indices for the first block.
    e) We report numerical results for $\alpha = 3$ with the geometry shown in d), and $p = 2$ Trotter formulas. In particular, we plot the expectation values of the stabilizer operators $O$ after the ancilla measurements, which should be $\langle O \rangle = \pm 1$ at the end of the surface code circuit.
    The results are averaged over $64$ initial Haar random states for the data qubits, and the dashed line is a guide for the eye following $\sim \tau^4$ (the $p = 2$ Trotter error squared). 
    f) Walsh sequences enable the implementation of such a quantum circuit in one-dimensional systems too, at the expense of using additional Walsh sequences to compensate for weakly-interacting links.
    We illustrate the realization of the first block with two Walsh sequences as an example, where the first one runs for a time $\tau_{1} = \tau$ and the second one for a larger time $\tau_2 = (3^{\alpha} - 1)\tau$.
    g) The numerical performances for the implementation of a one-dimensional system with $\alpha = 0.2$ are reported, where the error scaling is the same as the 2d case, and the values of the errors are similar.}
    \label{fig:surface_code}
\end{figure*}

As discussed in Sec.~\ref{sec:universal}, Walsh pulse sequences can find application in realizing quantum gates in analog quantum processors.
This can be done by effectively deleting interactions until qubits are interacting in pairs, realizing two-qubit gates.
Let us illustrate this powerful connection between quantum circuits and Walsh sequences in the context of quantum error correcting circuits. 

We consider the surface code, which is routinely realized with a set of physical qubits placed on a square lattice encoding the logical qubit, and ancilla qubits used to perform error detection~\cite{fowler_surface_2012}. 
The building block of the code is the measurement of stabilizer operators $\bigotimes_{i \in \text{plaq}(X^{(j)})} X_i$ ($\bigotimes_{i \in \text{plaq}(Z^{(j)})} Z_i$) involving the $4$ (or less) physical qubits connected to the ancilla qubit in the center, which is dubbed as $X^{(j)}$ ($Z^{(j)}$) (by abuse of notations). 
The standard circuit realizing the $X$ stabilizer measurement with CNOT gates is shown in Fig.~\ref{fig:surface_code}a). 
Such a circuit naturally maps to a sequence of two-qubit rotations parametrized by $[R_{XX}^{\pi/2}]_{ij}=\exp(iX_i X_j\pi/4)$ (see App.~\ref{app:surface_proof} for a quick derivation).
Note that these rotations can be realized using pair-wise Ising interactions.
In QEC, such measurements in the $X$ basis are complemented by measurements in another basis, which we choose here to be the $Y$ basis (instead of the traditional $Z$ basis). The circuit for $Y$ measurements is shown in Fig.~\ref{fig:surface_code}b). 

Let us assemble these different circuits blocks to show the implementation of a minimal surface code made of $4$ physical qubits, and $3$ ancilla bits, namely the $7-$surface code. 
The circuit realizing the $7-$surface code in terms of Ising interactions is shown in Fig.~\ref{fig:surface_code}c).
One realizes sequentially the three (commuting) measurements circuits involving two $X^{(i)}$ and one $Y$ ancilla qubits. 
In order to obtain an efficient mapping of Ising evolution that can be implemented withing our framework in parallel, we take advantage of reordering strategies that have been proposed for instance in the context of superconducting surface codes~\cite{versluis_scalable_2017}. 
This consists in permuting the order of commuting gates.
For the $7-$surface code, we then obtain the circuit in Fig.~\ref{fig:surface_code}c). 
Note that for larger codes, we can obtain in a similar way circuits that are also of depth $4$. 
In each of these $4$ layers shown in Fig.~\ref{fig:surface_code}c), every qubit is subject to at most one gate, thus we can implement all these gates in parallel.
In particular, within our framework, each layer corresponds to a different set of independent $XX$ and $YY$ Ising interactions.
We first consider an implementation of the $7-$surface code where the qubits have the same geometry as the ones in the surface code, and are interacting via power-law decaying interactions.
In this setting, each layer can be realized with a single Walsh sequence.
For instance, we show the Walsh sequence required to implement the first layer in Fig.~\ref{fig:surface_code}d).
We benchmark numerically this scenario for $\alpha = 3$ by plotting the expectation values of the stabilizer operators, which should be $\langle O \rangle = \pm 1$, following the ancilla measurement performed after the evolution with Walsh sequences.
The results, averaged over $64$ initial Haar random states for the data qubits, are reported in Fig.~\ref{fig:surface_code}e), where we observe errors decaying with $\sim \tau^4$, as expected for $p=2$ Trotter formulas, and an accuracy $|\langle O \rangle| \sim 1 - 10^{-6}$ already for $\tau / n \sim 10^{-2}$.

We also address the implementation of a $7-$surface code starting from a one-dimentional interacting qubit chain. 
In this geometry, certain layers might require $Q>1$ Walsh sequences.
This is because the interaction rescaling graph for each block has inhomogeneous weights.
For instance, as shown in Fig.~\ref{fig:surface_code}f), the first block has $g^X_{2,X^{(2)}} = 1$ and $g^Y_{1,Y} = 3^{\alpha}$.
Therefore, it needs two different Walsh functions, the first one implementing $g^X_{2,X^{(2)}} = 1$ and $g^Y_{1,Y} = 1$, for a time $\tau_{q = 1} = \tau$, and a second one implementing $g^X_{2,X^{(2)}} = 0$ and $g^Y_{1,Y} = 1$, for a time $\tau_{q=2} = (3^{\alpha} - 1)\tau$.
The total number of Walsh sequences needed to implement the surface code in this geometry is $Q = 5$.
In Fig.~\ref{fig:surface_code}g) we report numerical results for this implementation in a long-range interacting chain with $\alpha = 0.2$, mimicking a trapped-ion chain. 
We observe a very similar behaviour to the 2d case initially considered.
Note finally that our protocol can be naturally applied to more general quantum error correcting codes, such as the LDPC codes~\cite{Bravyi_2024}, which are raising recent interest. 

\section{Application to quantum optimization algorithms}
\label{sec:maxcut}

\begin{figure}[t]
    \includegraphics[width=0.47\textwidth]{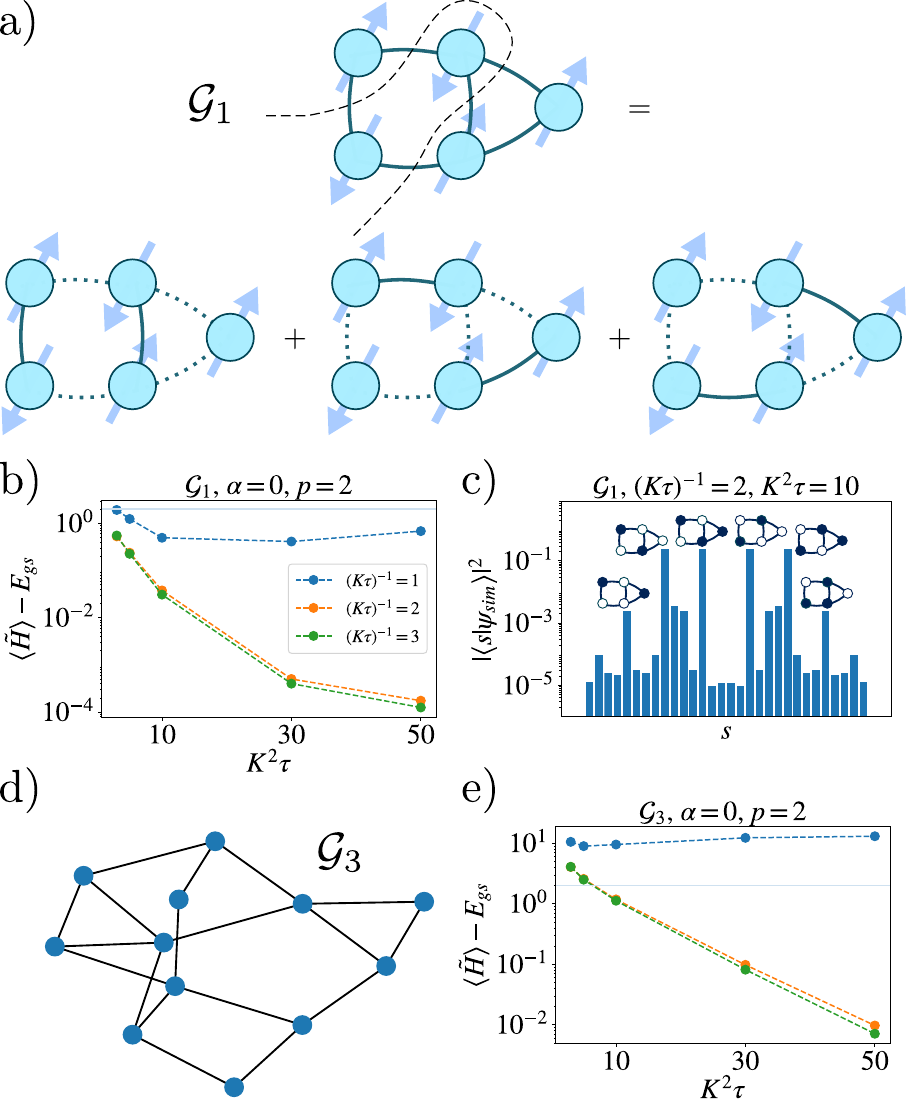}
    \caption{\textit{Digitized quantum annealing for the MaxCut problem with Walsh sequences} $-$ a) We show the degree 1 decomposition given by Algorithm~\ref{alg:graph_deco} for the graph $\mathcal{G}_1$, which decomposes it in $d_{\text{max}} = 3$ subgraphs, $d_{\text{max}}$ being the highest degree of $\mathcal{G}_1$. b) We plot the energy difference between the state prepared with the Walsh sequence dQA algorithm (Eq.~\eqref{eq:dQA_walsh}) and the ground state energy $E_{\text{gs}}$ of the MaxCut Hamiltonian $\tilde{H}$ (Eq.~\eqref{eq:maxcut_ham}). We observe that the discretization error related to $K\tau$ saturates, while increasing the annealing time improves exponentially the energy difference. c) We corroborate the quality of this approach plotting the amplitudes of the prepared state $\ket{\psi_{\text{sim}}}$, where $\ket{s}$ are spin configurations in the $x$ basis. d),e) We repeat the energy difference analysis for a larger graph $\mathcal{G}_3$, obtaining similar results. }
\label{fig:maxcut}
\end{figure}

While our first two examples could be implemented with typically few $O(1)$ Walsh sequences due to the low connectivity of $\tilde H$, let us address a final situation where one requires dense coupling matrices between qubits. 
This is the case for example in quantum optimization algorithms, where the solution to a complex computational problem is encoded in the ground of the Hamiltonian realized by a quantum simulator, see e.g.~\cite{pichler2018quantum, Ebadi_2022, Nguyen_2023, goswami2023solving} with Rydberg atoms.
In particular, we consider quantum optimization schemes with digital quantum annealing (dQA)~\cite{Barends_2016, mbeng2019quantum, hegade2021_dQA}.
dQA is a quantum algorithm that allows to approximate the ground state of a target Hamiltonian $\tilde{H}$ by starting from a product state which is the ground state $\ket{\psi_0}$ of a single qubit Hamiltonian $H_0$ such that $[\tilde{H}, H_0]\neq 0$, and evolving the system as
\begin{equation}
\label{eq:dQA_walsh}
    U_{dQA} = \prod_{k=1}^K
    e^{-i (K - k) \tau H_0}
    e^{-i\tau \tilde H k} \ .
\end{equation}
When $K\tau \ll 1$ this approach reduces to continuous-time quantum annealing, for which we know that, for large total annealing times $K^2\tau \gg 1$, (with requirements related to the spectral gap of the instantaneous Hamiltonian~\cite{Zhou_2020_qaoa}) the dynamics is adiabatic and we can prepare in good approximation the ground state of $\tilde{H}$.

We demonstrate dQA with Walsh functions considering the MaxCut problem that consists in finding the ground-state of an Ising Hamiltonian
\begin{equation}
\label{eq:maxcut_ham}
    \tilde{H} = J\sum_{i<j} G_{ij}X_i X_j
\end{equation}
where $G_{ij}$ is the adjacency matrix of the target graph $\mathcal{G}$.
For simplicity, we consider, as a resource Hamiltonian, Eq.~\eqref{eq:LR_coupling} with $\alpha = 0$, meaning $J_{ij} = J$ $\forall i\neq j$.
In this case, the interaction rescaling graph for the $XX$ interaction is exactly $g^X_{ij} = JG_{ij}/(-J) = -G_{ij}$, while for the $YY$ interaction $g^Y_{ij} = 0$.
Therefore, to implement $\tilde{U}(\tau)$, we first decompose $\mathcal{G}$ in degree 1 graphs using Algorithm~\ref{alg:graph_deco}.
Then, for each degree 1 graph $\mathcal{G}_q'$ we design a Walsh sequence implementing it for the $XX$ interaction, while we decouple all the $YY$ interactions by choosing $y_i\neq y_j$ $\forall i\neq j$.
To flip the sign of the resulting Hamiltonian, it is sufficient to apply the pulse $P_{\text{set}}^q$ before and after each Walsh sequence $
    P_{\text{set}}^q = \bigotimes_{i, (i, j)\in E(\mathcal{G}_q')}Y_i$, which flips the sign of one of the two $X$ operators on each link of the graph.
We choose the initial state to be $\ket{\psi_0} = \bigotimes_i \ket{0}_i$, and therefore $H_0 = -\sum_i Z_i$. The evolution under $H_0$ can be implemented with global pulses.

We numerically benchmark this protocol by computing the energy difference $\langle\tilde{H}\rangle - E_{\text{gs}}$ between the prepared state and the ground state of $\tilde{H}$ as a function of the discretization time $K\tau$ and the total annealing time $K^2\tau$.
The numerical results are reported for the two graphs $\mathcal{G}_1$ and $\mathcal{G}_3$ in Fig.~\ref{fig:maxcut}.
It is possible to see that the discretization error (due to $K\tau > 0$) quickly saturates after $(K\tau)^{-1} \ge 2$, and so in that regime only increasing the total annealing time $K^2 \tau$ reduces the energy difference.
In particular, the energy difference reduces exponentially with $K^2 \tau$.
To further validate our method, we plot the amplitudes of the prepared state $\ket{\psi_{\text{sim}}}$ in the $x$ basis, which give us the probability of a certain bitstring $s$ as measurement outcome.
For parameters giving a energy difference $\langle \tilde{H} \rangle - E_{\text{gs}} \simeq 10^{-2}$, we observe that the `wrong' solutions for the MaxCut problem have at most a probability $|\langle s | \psi_{\text{sim}}\rangle|^2 \sim 10^{-3}$, which is two order of magnitude less likely than the correct outcomes.
Note that we can straightforwardly adapt the present example to different quantum optimization circuits, such as the quantum approximate optimization algorithm (QAOA)~\cite{farhi2014quantum, Zhou_2020_qaoa}.

\section{Implementation }
\label{sec:experiments}
Finally, we briefly comment on the experimental realizations of our protocol. While we believe our protocol can also be easily implemented with solid-state qubits, we focus here for concreteness on atomic qubits.

\subsection{Rydberg atoms}
\label{subsec:rydbergs}
Rydberg atoms trapped in optical tweezer arrays~\cite{Barredo2016,Endres2016,browaeys2020review} provide a natural experimental platform for the implementation of the proposed 
framework to universal quantum computing and simulation, featuring long-range interactions as well as fast, local pulse operations.
The first proposed implementation builds on the natural emergence of the resource Hamiltonian $H_R$ in the interaction between two opposite-parity Rydberg states, which interact via resonant dipole-dipole interactions.
These interactions have been utilized for quantum simulations of the $XY$-model using tweezer-trapped atoms~\cite{barredo_coherent_2015, DeLeseleuc2019,scholl_microwave_2022,chen2023continuous}.
The required local single-qubit gates can be constructed in the Rydberg-qubit basis by a combination of two global microwave rotations $R_X^{\pi/2}$ about the X-axis by an angle of $\pi/2$ combined with three local phase shifts, which correspond to rotations about the Z-axis, $R_Z^{\theta_i}$, as~\cite{Wang_2016_weiss}
\begin{align}
R_{O}^{\theta} = R_Z^{\theta_1} R_X^{\pi/2} R_Z^{\theta_2} R_X^{-\pi/2} R_Z^{\theta_3}.
\end{align}
This Euler decomposition~\cite{McKay2017,Nottingham2023a} of an arbitrary single-qubit rotation $R_{O}^{\theta}$ has been realized experimentally in a chain of trapped ions~\cite{Brydges2019}.
The required local $\pi$-pulses about the $X$-axis ($Y$-axis) are given by the choice $(\theta_1, \theta_2, \theta_3)=(\pi/2, \pi, -\pi/2)$ ($(\theta_1, \theta_2, \theta_3)=(-\pi, \pi, \pi)$) respectively.
Local phase rotations by controllable angles can be experimentally implemented through spatial addressing of individual atoms with acousto-optic deflectors, using off-resonant coupling to lower-lying states~\cite{Notarnicola2023}. 
This induces light-shifts selectively on one of the two dipole-coupled Rydberg states, leading to the required phase rotations in the qubit basis.
In experiments, typical interaction strengths can reach up to $1\,$MHz at distances of approximately $10\,\mu$m~\cite{chen2023continuous}.
We note that our pulse schemes are robust against small variations in the interaction strengths of approximately $2\%$ arising from the finite spread of typically $30\,$nm of atoms trapped in the motional ground state of typical optical tweezer or lattice potentials.
To suppress finite-pulse-duration errors, see Sec.~\ref{subsec:finite_pulses}, the single-qubit manipulations must be significantly faster than the timescale set by the interaction strength in $H_R$.
Experimentally, this criterion can be fulfilled, with both light-shifts and also global microwave couplings exceeding $10\,$MHz in state-of-the-art experiments~\cite{chen2023continuous}, with a straightforward scaling perspective of at least one order of magnitude faster pulses.
The associated timescales with interaction and single qubit gates are significantly shorter than typical Rydberg state lifetimes on the order of $100\,\mu$s, which can be significantly increased e.g., using circular Rydberg states~\cite{Wu2023, Ravon2023,Holzl2024}.

In an alternative approach, the qubits can be encoded in atomic ground or low-lying metastable states.
Here, single-qubit gates can be performed using single- or two-photon coupling with coupling strengths in the MHz-regime~\cite{bluvstein_quantum_2022}.
The resource Hamiltonian can be generated using off-resonant Rydberg dressing, where the Rydberg character is optically admixed to the ground state.
Rydberg-dressed interactions have been realized and benchmarked in optical lattices~\cite{Zeiher2016,Zeiher2017,Schine2022,Eckner2023}, optical tweezers~\cite{Jau2016,Steinert2023} and bulk gases~\cite{Hines2023}.
The optical switchability of the interactions in this case allows for a direct realization of the required pulse-schemes with alternating global application of the resource Hamiltonian and single-qubit rotations, thereby eliminating the need to compensate for finite pulse durations.
Furthermore, controlling the dressed admixture, the timescale set by the resource Hamiltonian can be tuned, such that also more complex single-qubit pulse sequences can be engineered.
During the application of the single-qubit gates, the resource Hamiltonian can also be turned off, such that the effective decoherence due to finite Rydberg lifetimes is minimized.
To generate the required $XY$-type interactions in $H_R$, two-color dressing schemes have been proposed and experimentally implemented~\cite{Glaetzle2015,VanBijnen2015,dlaska2017dlaska,Steinert2023}.

\subsection{Trapped ions}
\label{subsec:trapped_ions}
With trapped ions, the resource Hamiltonian amounts to a long-range Ising model ($J^Y_{ij}=0$)~\cite{monroe2021programmable}, where each ion encodes a spin using two long-lived electronic states. 
These interactions decay as a power-law with tunable exponent $\alpha$.
Since $H_R$ is here laser-mediated, the time-evolution can be interspersed efficiently with single-qubit pulses~\cite{rajabi_dynamical_2019,kranzl_experimental_2023,morong_engineering_2023}.
Finally, these pulses can be programmed in a local way using experimentally-demonstrated optical addressing techniques, see e.g. Ref.~\cite{Brydges2019}.

\subsection{Polar molecules}
\label{subsec:molecules}
Arrays of ultracold polar molecules have been considered for a long time as candidates to perform quantum processing tasks~\cite{DeMille_2002, Karra_2016}.
Molecules feature rich internal structure, long coherence times~\cite{Park2017b,Gregory2024}, and long-range dipolar interactions. Recently, single molecules have been directly laser cooled~\cite{Anderegg2019} 
or associated from single atoms~\cite{Liu2018} before trapping in tweezers. Recently, such tweezer-trapped single molecules also have been entangled successfully via dipolar interactions~\cite{Holland2023,Bao2023}.
We argue that ultracold polar molecules trapped in optical tweezers or lattices can be used to realize universal quantum processing with Walsh pulse sequences.
In particular, it is possible to encode a spin in the rotational states.
Together with the dipolar interactions, we obtain a resource Hamiltonian similar to Eq.~\eqref{eq:H_resource}. The spin coupling depends strongly on the exact implementation and molecular species, with values as of $J^X_{ij} = J^Y_{ij} \sim 2\pi \times 210 \text{ Hz}$ realized recently in an optical lattice of spacing $a\sim 540 \text{ nm}$~\cite{miller2024twoaxis} or $J^X_{ij} = J^Y_{ij} \sim 2\pi \times 43 \text{ Hz}$ in optical tweezers at $a\sim1.93 \,\mu$m~\cite{Holland2023}.
In lattices, Hamiltonian engineering with global pulse sequences has been already demonstrated~\cite{miller2024twoaxis}.
The use of rotational states implies the need of global microwave fields in the GHz-range to perform $X$ pulses.
The experimental toolbox also allows for applying local AC Stark shifts for local molecular control away from magic trapping conditions for the qubits.
This indicates that the building blocks to implement Walsh pulse sequences in array of polar molecules are currently within reach.

\section{Conclusions and outlook}

This work proposes a complete and robust framework for building universal quantum processors in quantum spin systems with existing quantum technology. 
This builds on the use of Walsh functions to parametrize in a flexible and robust way pair-wise interactions in average Hamiltonian theory.
Our work points to future interesting directions. 

First, while we can effectively switch on/off interactions of the resource Hamiltonian, we cannot with the present proposal increase its connectivity. Thus, it will be interesting to combine our approach with strategies based on moving atoms~\cite{bluvstein_quantum_2022} to achieve programmable connectivity in a robust and time-optimal fashion. An alternative strategy would be to use ancilla qubits realizing, withing our framework, multiple SWAP gates, to effectively enhance the connectivity.

Furthermore, the possibility to `isolate' interacting groups of spins can find further applications beyond Hamiltonian engineering. 
In particular, this can be used for robust Hamiltonian learning in which interacting pairs are isolated~\cite{huang_learning_2023}, 
and realizations and couplings of identical quantum systems for entanglement measurements~\cite{horodecki2003measuring,carteret2005noiseless,gray2018machine,bluvstein_quantum_2022}.
Our protocol should also find extensions beyond qubits, in particular for fermionic/bosonic atoms in optical lattices~\cite{schafer2020tools}. In this case, a Hubbard model evolution can be used as resource Hamiltonian, and the dynamics can be enriched with local pulses (Stark-shifts or beam splitter operations) that can be described in AHT. It would be interesting in particular to see if Walsh functions can be used to build fermionic quantum processors~\cite{gonzales2023fermionic} from analog systems with dipolar interactions, such as fermionic magnetic atoms~\cite{chomaz2022dipolar}.

\begin{acknowledgements}
We thank J. Ignacio Cirac, Dolev Bluvstein and Shannon Whitlock for insightful discussions.
Work in Grenoble is funded by the French National Research Agency via the JCJC project QRand (ANR-20-CE47-0005), and via the research programs Plan France 2030 EPIQ (ANR-22-PETQ-0007), QUBITAF (ANR-22-PETQ-0004) and HQI (ANR-22-PNCQ-0002).
J.Z. acknowledges funding by the Max Planck Society (MPG) the Deutsche Forschungsgemeinschaft (DFG, German Research Foundation) under Germany's Excellence Strategy--EXC-2111--390814868, and from the Munich Quantum Valley initiative as part of the High-Tech Agenda Plus of the Bavarian State Government.
This publication has also received funding under Horizon Europe programme HORIZON-CL4-2022-QUANTUM-02-SGA via the project 101113690 (PASQuanS2.1).
J.Z. also acknowledges support from the BMBF through the program “Quantum technologies - from basic research to market” (Grant No. 13N16265).
\end{acknowledgements}

\bibliographystyle{quantum}
\bibliography{walsh} 

\newpage
\onecolumn
\appendix

\section{Walsh functions}
\label{app:walsh}

A central tool in the design of our protocol are Walsh functions.
Walsh functions are an orthonormal family of binary and discrete functions of time, introduced in harmonic analysis~\cite{walsh1923closed, chien1975representations_walsh} and became popular in signal theory~\cite{beer1981walsh}.
In our work, we use a specific representation of Walsh functions, which relates them to Hadamard matrices.
Concretely, we define the unnormalized Hadamard matrix as
\begin{equation}
H_1 = 
\begin{pmatrix}
    1 & 1 \\
    1 & -1
\end{pmatrix} 
\end{equation}
and, recursively, a family of generalized Hadamard matrices as 
\begin{equation}
H_q = H_1 \otimes H_{q-1} = 
\begin{pmatrix}
    H_{q-1} & H_{q-1} \\
    H_{q-1} & -H_{q-1}
\end{pmatrix} 
\end{equation}
where we intend the tensor product as Kronecker product.
We then define Walsh functions from the rows of a generalized Hadamard matrix $H_q$, i.e.
\begin{equation}
    w^{(k)}_a = [H_q]_{a,k}
\end{equation}
with $a = 0, 1, ..., 2^q - 1$.
Since $a < 2^{q}$, in order to choose high Walsh indices $a$ we need to increase $q$ accordingly.
From this, we obtain that the periodicity of a Walsh function is
\begin{equation}
    n = 2^q,\ q = \lceil \log_2 (a) \rceil
\end{equation}
which translates to the length of pulse sequences designed with Walsh functions in Eq.~\eqref{eq:length_pulse_seq}.
The orthonormality of Walsh functions is then guaranteed by the unitarity and the symmetry of the Hadamard matrices: since $[H_q^2]_{ij} = n\delta_{ij}$, and $H_q^T = H_q$, then
\begin{equation}
    (w_a|w_b) := \frac{1}{n}\sum_k w^{(k)}_a w^{(k)}_b = \delta_{ab} \ .
\end{equation}
A straightforward property used multiple times in the text can be now derived: since $w^{(k)}_0 = 1$, we obtain
\begin{equation}
    \frac{1}{n}\sum_k w^{(k)}_a = \frac{1}{n}\sum_k w^{(k)}_0 w^{(k)}_a = (w_0|w_a) = \delta_{0a} \ .
\end{equation}

\section{Proof of universality for two-body Hamiltonians}
\label{app:proof_universality}

In this section, we prove that any two-body spin Hamiltonian can be realized in AHT using multiple Walsh sequences.
Let us assume at first that we want to prove the possibility of engineering any possible Hamiltonian of the form
\begin{equation}
\label{eq:app_univ_Hamiltonian}
        \tilde H = \sum_{i<j} (\tilde{J}^X_{ij}X_iX_j + \tilde{J}^Y_{ij}Y_iY_j) 
\end{equation} 
using Eq.~\eqref{eq:H_resource} as resource Hamiltonian and multiple Walsh sequences.
We recall that a single Walsh sequence provides the average Hamiltonian
\begin{equation}
    \tilde H^{(q)} = \sum_{i<j} \left([G^X_q]_{ij}J^X_{ij}X_iX_j + [G^Y_q]_{ij}J^Y_{ij}Y_iY_j\right)
\end{equation}
where $[G^X_q]_{ij}$ ($[G^Y_q]_{ij}$) is the adjacency matrix of the graph $\mathcal{G}^X_q$ ($\mathcal{G}^Y_q$), the graphs being defined by
\begin{equation}
\label{eq:app_graph}
    \mathcal{G}^X_q = \{(i,j)\ |\ x^q_i=x^q_j\},\ \mathcal{G}^Y_q = \{(i,j)\ |\ y^q_i=y^q_j\} 
\end{equation}
with $x_i^q$ and $y_i^q$ the Walsh indices chosen in the $q-$th Walsh sequence.
Since the two graphs $\mathcal{G}^X_q$ and $\mathcal{G}^Y_q$ can be parametrized independently, we can focus on the $XX$ part of the Hamiltonian. 
Alternating $Q$ Walsh sequences, each one having a different duration $\tau_q$, realizes a unitary dynamics $\tilde{U} (\tau) = e^{-i\tau\tilde{H}} + O(\tau^{p+1})$, where
\begin{equation}
    \tilde{J}^X_{ij} = \frac{1}{\tau}J^X_{ij}\sum_q \tau_q [G^X_q]_{ij} \ .
\end{equation}
We then define the interaction rescaling as
\begin{equation}
\label{eq:app_int_rescaling}
    g_{ij} := \tilde{J}^X_{ij} / J^X_{ij} = \sum_q c_q [G^X_q]_{ij} 
\end{equation}
where $c_q := \tau_q/\tau$, and we assume here $J_{ij}\neq 0$.
Seeing $g_{ij}$ as the adjacency matrix of a weighted graph $\mathcal{G}$, we note that this definition amounts to a weighted decomposition of $\mathcal{G}$ in subgraphs $\mathcal{G}^X_q$ respecting Eq.~\eqref{eq:app_graph}.
Therefore, to prove that we are able to engineer any Hamiltonian of the form of Eq.~\eqref{eq:app_univ_Hamiltonian} amounts to prove that $\mathcal{G}$ can be any graph.
The proof is done in three steps of increasing complexity:
\begin{enumerate}
    \item if $g_{ij}\in \{0, 1\}$, then we can take $c_q = 1\ \forall q$ in Eq.~\eqref{eq:app_int_rescaling}, and prove that any unweighted graph $\mathcal{G}$ can be decomposed in subgraphs respecting Eq.~\eqref{eq:app_graph};
    \item if $g_{ij}>0$ with more general values, it is possible to extend the above proof by decomposing each weighted subgraph in a set of unweighted subgraphs with coefficients $c_q>0$;
    \item if $g_{ij}$ can also be negative, we know from Eq.~\eqref{eq:app_int_rescaling} that multiple Walsh sequences are not enough to prove universality ($c_q>0$ by definition), but we will prove that it is sufficient the introduction of few single-qubit pulses in between Walsh sequences.
\end{enumerate}
Finally, we will show that the introduction of single-qubit pulses in between Walsh sequences allows us to engineer two-body interactions between arbitrary spin operators. 

\subsection{Case I: $g_{ij}\in\{0, 1\}$}
We first consider the case in which $g_{ij}\in \{0, 1\}$.
For power-law decaying couplings described by Eq.~\eqref{eq:LR_coupling}, this could represent the case of engineering a model with only nearest-neighbor couplings with coupling strength strength $J$, such as in Sec.~\ref{sec:ising_chain}.
Further couplings are admitted only if their coupling strength is $J/r_{ij}^{\alpha}$.
Another relevant case is the one in which the resource Hamiltonian has $\alpha = 0$, and the target interaction is $J_{ij} = G_{ij}J$, i.e. the coupling strength is homogeneous and solely defined by a graph structure, such as in Sec.~\ref{sec:maxcut}.
In all of these cases, we take $\tau_q = \tau \ \forall q$, obtaining 
\begin{equation}
    g_{ij} = \sum_q (G^X_q)_{ij} \ .
\end{equation}
Since $g_{ij}\in\{0, 1\}$ can be seen the adjacency matrix of a target unweighted graph $\mathcal{G}$, this equation amounts to a graph decomposition in terms of graph respecting Eq.~\eqref{eq:app_graph}.
In particular, graphs of degree $1$ fulfill this condition, using $x_i=x_j$ only for the connected pairs. As any graph can be decomposed in graphs of degree $1$, the decomposition of $g_{ij}$ is always possible. 
One can proceed in different ways:
\begin{enumerate}
    \item We can always decompose a graph $\mathcal{G}(V, E)$ in terms of all its links $E$
    \begin{equation}
        g_{ij} = \sum_{(k, l)\in E} \delta_{k, l} g_{ij}
    \end{equation}
    where each degree 1 graph corresponds to a single link. This approach is however inefficient, since a densely connected graph can decomposed in up to $O(N^2)$ subgraphs;
    \item Any complete graph $K_N$ (with even $N$) can be decomposed in terms of $N/2$ open (Hamilton) paths $P_N^q$~\cite{Galicia_2020_connectivity, Bryant_2013, shyu2010decomposition} with $N^2/2$ assignments.
    In turn, any open path $\mathcal{P}_N^q$ can be decomposed in two degree 1 subgraphs $\mathcal{P}_N^q = \mathcal{G}_q' + \mathcal{G}_q''$ by alternating links along itself ($\mathcal{G}_q'$ consisting of the odd links, $\mathcal{G}_q''$ consisting of the even links). Having decomposed any complete graph in terms of degree 1 graphs, we have achieved the decomposition of an arbitrary graph as well
    \begin{equation}
        g_{ij} = \sum_{q} [G_q']_{ij}g_{ij} + \sum_{q} [G_q'']_{ij}g_{ij} \ .
    \end{equation}
    Note that this decomposition results in at most $N$ subgraphs, which is optimal for complete graphs, but not for all graphs. Additionally, its runtime $O(N^2)$, is comparable with listing all links (see~\cite{Galicia_2020_connectivity} for the explicit algorithm). For $N$ odd, a similar construction holds, resulting in at most $N + O(1)$ subgraphs;
    \item Algorithm~\ref{alg:graph_deco} achieves a decomposition of any graph $\mathcal{G}$ (of maximum degree $k_{\text{max}}$) in a set of $Q = O(k_{\text{max}})$ graphs of degree 1, i.e.
    \begin{equation}
        g_{ij} = \sum_{q} [g_q^1]_{ij} \ .
    \end{equation}
    This algorithm has a runtime of order $O(N^3 k_{\text{max}})$. While classically it takes more time than the previous approach, the number of resulting subgraphs is smaller than $N$ for a large number of graphs (according to their sparsity). This is less expensive for the resulting quantum simulation, requiring a smaller number of Walsh sequences to engineer the same Hamiltonian. For instance, we apply this method in Sec.~\ref{sec:maxcut}, where we consider the realization of Hamiltonians with complex interaction graphs.
\end{enumerate}

\begin{algorithm}[t]
\caption{Greedy algorithm for the decomposition of $\mathcal{G}$ in degree 1 graphs $\mathcal{G}'_q$}
\SetKwInOut{Input}{Target graph}
\Input{$\mathcal{G}$}
$\mathcal{G}_r\gets \mathcal{G}$\;
$q\gets 1$\;
\While{$E(\mathcal{G}_r)\neq\emptyset$}{
\textbf{Def:} $\mathcal{G}_q'$\;
$\mathcal{G}_q'\gets \mathcal{G}$\;
$E(\mathcal{G}_q')\gets\emptyset$\;
\While{$V(\mathcal{G}_r)\neq\emptyset$}{
\textbf{pick} $i\in V(\mathcal{G}_r)$ with max degree\;
\textbf{pick} $j\in V(\mathcal{G}_r)$, such that $(i, j)\in E(\mathcal{G}_r)$, with max degree\;
$E(\mathcal{G}_q')\gets E(\mathcal{G}_q') + (i, j)$\;
$V(\mathcal{G}_r) \gets V(\mathcal{G}_r) - i - j$\;
}
$E(\mathcal{G}_r) \gets E(\mathcal{G}_r) - E(\mathcal{G}_q')$\;
$q \gets q+1$
}
\label{alg:graph_deco}
\end{algorithm}

\subsection{Case II: $g_{ij}>0$}

We now consider the case in which the target Hamiltonian has a different geometry than the resource Hamiltonian (i.e. it requires $g_{ij}\in \mathbb{R}$).
For the moment, we restrict to the case in which $g_{ij}>0$.
We want to prove that any positive-weighted adjacency matrix can be written as
\begin{equation}
\label{eq:app_weighted_deco}
    g_{ij} = \sum_q \frac{\tau_q }{\tau}[G^X_q]_{ij}
\end{equation}
where $(G^X_q)_{ij}$ are unweighted adjacency matrices of graphs $\mathcal{G}^X_q$ of degree 1.
As a first step, following one of the procedures given in the previous subsection, we decompose the graph $\mathcal{G}$ in terms of inhomogeneously weighted degree 1 graphs
\begin{equation}
\label{eq:app_unweighted_deco}
    g_{ij} = \sum_p [g_p']_{ij} 
\end{equation}
where the weights of the links $E_p$ in the subgraphs are the same as the ones of the links in $E$.
Then, for each degree 1 graph $\mathcal{G}_p'$, we call the smallest weight $c_{(p, 1)}$, and we define the decomposition
\begin{equation}
     [g_p']_{ij}^{(1)} = [g_p']_{ij} - c_{(p, 1)}[g_{(p, 1)}'']_{ij}
\end{equation}
where $\mathcal{G}_{p, 1}''$ is an unweighted graph with the same topology as $\mathcal{G}_p'$.
Iterating this decomposition on the resulting graph $[g_p']_{ij}^{(1)}$ for all the values of the weights, we obtain
\begin{equation}
\label{eq:app_deg1indeg1_deco}
    [g_p']_{ij} = \sum_k c_{(p, k)} [g_{(p, k)}'']_{ij} 
\end{equation}
with $c_{(p, k)} > 0$. Plugging this equation in Eq.~\eqref{eq:app_weighted_deco}, we obtain
\begin{equation}
    g_{ij} = \sum_{(p, k)} c_{(p, k)} [g_{(p, k)}'']_{ij} \ .
\end{equation}
Taking $(p, k) = q$, and $c_q = \tau_q / \tau$, we have shown that it is possible to reach the decomposition of Eq.~\eqref{eq:app_unweighted_deco}, where $\{\mathcal{G}_q^X\}$ are graphs of degree 1.
We can follow this same procedure when designing pulse sequences: for each weighted graph of degree 1, we will require as many Walsh sequences as there are different weights.
While this case requires more Walsh sequences than the one of the previous subsection, following the decomposition of Eq.~\eqref{eq:app_deg1indeg1_deco} will require a total time $\sum_k\tau_{(p, k)} = \tau \cdot \text{max}_{(i, j)\in E_p}(g_{ij})$, like in the latter case.

\subsection{Case III: arbitrary $g_{ij}$ and two-body Hamiltonians}

Having demonstrated how to realize any possible $g_{ij}>0$ using Walsh sequences, we now address the more general case in which we can also have $g_{ij}<0$.
From the previous proof, if each degree 1 subgraph we obtain in  Eq.~\eqref{eq:app_weighted_deco} can take values $[g^X_q]_{ij}\in \{-1, 0, 1\}$, then $g_{ij}$ can take any possible real value.
We now show that, to obtain $[g^X_q]_{ij}\in \{-1, 0, 1\}$ it is not necessary to add more Walsh sequences, but just a single pulse at the beginning and at the end of the $q-$th Walsh sequence.
To begin with, we recall that the $q-$th Walsh sequence realizes the product formula $\tilde{U}^{(q)}(\tau) \simeq e^{-i\tau{H}^{(q)}}$, with average Hamiltonian
\begin{equation}
    {H}^{(q)} = \sum_{i<j}( |[g^X_q]_{ij}| J_{ij} X_i X_j + |[g^Y_q]_{ij}| J_{ij} Y_i Y_j ) = \sum_{i<j}( [\hat{h}^{XX}_q]_{ij} + [\hat{h}^{YY}_q]_{ij} ) \ .
\end{equation}
Assuming that the graph $\mathcal{G}^X_q + \mathcal{G}^Y_q$ is also of degree 1 (if it is not the case, it can be decomposed in two degree 1 graphs), the product formula can be written as
\begin{equation}
    \tilde{U}^{(q)}(\tau) \simeq \bigotimes_{(i, j)\in E^X,\ (k, l)\in E^Y} e^{-i\tau[\hat{h}^{XX}_q]_{ij}} e^{-i\tau[\hat{h}^{YY}_q]_{kl}} \ .
\end{equation}
We now divide the links in two subsets $E^X = E^X_+ + E^X_-$, where $(i, j) \in E^X_+$ ($(i, j) \in E^X_-$) if $[g^X_q]_{ij}>0$ ($[g^X_q]_{ij}<0$). 
We define the setting pulse $P^{\text{set}} = \bigotimes_i p^{\text{set}}_i$, where:
\begin{enumerate}
    \item $p^{\text{set}}_i = X$ if $(i, j)\notin E^X_-$ and $(i, j)\in E^Y_-$;
    \item $p^{\text{set}}_i = Y$ if $(i, j)\in E^X_-$ and $(i, j)\notin E^Y_-$;
    \item $p^{\text{set}}_i = Z$ if $(i, j)\in E^X_-$ and $(i, j)\in E^Y_-$;
    \item $p^{\text{set}}_i = 1$ otherwise.
\end{enumerate}
Let us stress that the assumption that the underlying subgraph $\mathcal{G}^X_q + \mathcal{G}^Y_q$ is of degree 1 is crucial, as it allows us to define the `set' pulse sequence in terms of individual links.
Recalling that 
\begin{equation}
    Z_i e^{-iJ\tau X_i X_j} Z_i =  e^{-iJ\tau (Z_i X_i Z_i) X_j} = e^{+iJ\tau X_i X_j} 
\end{equation}
applying the setting pulse results in the effective unitary evolution
\begin{equation}
    P^{\text{set}}\tilde{U}^{(q)}(\tau)P^{\text{set}} \simeq \bigotimes_{(i, j)\in E^X_+,\ (k, l)\in E^Y_+} e^{-i\tau[\hat{h}^{XX}_q]_{ij}} e^{-i\tau[\hat{h}^{YY}_q]_{kl}} \bigotimes_{(i', j')\in E^X_-,\ (k', l')\in E^Y_-} e^{i\tau[\hat{h}^{XX}_q]_{i'j'}} e^{i\tau[\hat{h}^{YY}_q]_{k'l'}} = e^{-i\tau\tilde{H}^{(q)}}\ .
\end{equation}
where the new average Hamiltonian is now our target Hamiltonian
\begin{equation}
    \tilde{H}^{(q)} = \sum_{i<j}( [g^X_q]_{ij} J_{ij} X_i X_j + [g^Y_q]_{ij} J_{ij} Y_i Y_j )  \ .
\end{equation}
At this point, we have proven the possibility to engineer, with $Q$ Walsh sequences and at most $Q+1$ additional pulses (which can be absorbed into the Walsh sequences anyway) any Hamiltonian of the form Eq.~\eqref{eq:app_univ_Hamiltonian}.
This proof can be extended by considering the case in which our target Hamiltonian is a general two-body spin Hamiltonian of the form
\begin{equation}
    \tilde{H} = \sum_{m=1}^{M}\sum_{i<j} J_{ij}^{(m)} O_i^{(m, 1)} O_j^{(m, 2)} 
\end{equation}
where $O_i^{(m, 1)}$ and $O_i^{(m, 2)}$ are two sets of spin operators (Pauli operators or rotations of Pauli operators).
To achieve this result, we repeat the construction of this subsection, but splitting the links in $m$ subsets $E^X = \sum_m E_m^X$.
Then, for $(i, j)\in E_m$, we can choose $p_i^{\text{set}}$ and $p_j^{\text{set}}$ such that
\begin{equation}
    p_i^{\text{set}} X_i (p_i^{\text{set}})^{-1} = O_i^{(m, 1)},\ p_j^{\text{set}} X_j (p_j^{\text{set}})^{-1} = O_j^{(m, 2)} \ .
\end{equation}
For instance, for $O_i^{(m,1)}=Z_i$, we apply the Hadamard gate $p_i^{\text{set}}=H$.
This gives
\begin{equation}
    (i, j)\in E^X_m \rightarrow p_i^{\text{set}}p_j^{\text{set}} e^{-iJ\tau X_i X_j} (p_i^{\text{set}}p_j^{\text{set}})^{-1} = e^{-iJ\tau O_i^{(m, 1)} O_j^{(m, 2)}} \ .
\end{equation}
Note that for $M = 2$, $O^{(m, 1)} = -X$, $O^{(m, 2)} = X$, and similarly the $YY$ part, the generalization reduces to the previous case.
We have thus proven that, with the same resources as for engineering arbitrary Hamiltonians as in Eq.~\eqref{eq:app_univ_Hamiltonian}, it is possible to engineer arbitrary two-body spin Hamiltonians.
Note that this result implies the possibility to realize universal quantum computing, for instance using as gate set arbitrary phase gates $\exp(-i\theta X_iX_j)$ and single qubit rotations.

\section{Trotter errors in Walsh sequences}
\label{app:errors}

We discuss Trotter errors obtained by engineering a Hamiltonian $\tilde{H}$ using a pulse sequence designed with Walsh functions.
In a general $p = 1$ Trotter formula, to realize a Hamiltonian $\tilde{H}$ for a time $T$, we run a set of Hamiltonians $\{H^{(k)}\}$ each one for a time $\tau_k$, with $\frac{T}{\sum_k \tau_k}$ cycles and $\tilde{H} = \frac{\sum_k \tau_k H^{(k)}}{\sum_k \tau_k}$.
The Trotter errors for such a general setting can be upper bounded, at order $O\left(\left(\sum_k\tau_k\right) T \right)$, as~\cite{lloyd_universal_1996}
\begin{equation}
\label{eq:app_general_lloyd}
    \left|\left|\left(\tilde{U}\left(\sum_k\tau_k \right)\right)^{T / \sum_k\tau_k} -  e^{-iT\tilde{H}}\right|\right| \leq \frac{1}{2}\frac{T}{\sum_k \tau_k} \Big{|} \Big{|} \sum_{k<l} [\tau_k H^{(k)}, \tau_l H^{(l)}] \Big{|} \Big{|} + O\left(\left(\sum_k\tau_k \right)^2 T \right)
\end{equation}
which reduces to Eq.~\eqref{eq:lloyd_formula} for a single Walsh sequence ($\tau_k = \tau / n$).
We consider the spectral norm $||O||_{\infty}:= \text{max}_{v} \frac{||Ov||}{||v||}$ (which for finite Hilbert spaces coincides with the largest singular value of the operator $O$), which is 1 for the identity and any Pauli operator.

At first we focus on the $Q = 1$ case, in which the average Hamiltonian $\tilde{H}$ is obtained from a single Walsh sequence, which reduces to Eq.~\eqref{eq:lloyd_formula}. be simplified through the following upper bound
\begin{equation}
    \mathcal{E}(\tau, T) := ||(\tilde{U}(\tau))^{T / \tau} -  e^{-iT\tilde{H}}|| = \frac{\tau T}{2n^2} \Big{|} \Big{|} \sum_{k<l} [H^{(k)}, H^{(l)}] \Big{|} \Big{|} + O(\tau^2 T) \leq \frac{\tau T}{2n^2}  \sum_{k<l} ||[H^{(k)}, H^{(l)}]|| + O(\tau^2 T) \ .
\end{equation}
The toggling-frame Hamiltonians obtained in applying Walsh sequences to these systems are of the form
\begin{equation}
    H^{(k)} = \sum_{i<j}( J_{ij}^X  w_{x_i}^{(k)}w_{x_j}^{(k)} X_i X_j + J_{ij}^Y w_{y_i}^{(k)} w_{y_j}^{(k)} Y_i Y_j)
\end{equation}
where $w_{x_i}^{(k)} = \pm 1$, $w_{y_i}^{(k)} = \pm 1$.
The commutator between two toggling-frame Hamiltonians is
\begin{eqnarray}
    \nonumber [H^{(k)}, H^{(l)}] & = \sum_{i<j,\ i'<j'} ( J_{ij}^X J_{i'j'}^Y w_{x_i}^{(k)}  w_{x_j}^{(k)}  w_{y_{i'}}^{(l)}  w_{y_{j'}}^{(l)} [X_i X_j, Y_{i'}Y_{j'}] +  J_{ij}^Y J_{i'j'}^X w_{y_i}^{(k)}  w_{y_j}^{(k)}  w_{x_{i'}}^{(l)}  w_{x_{j'}}^{(l)} [Y_i Y_j, X_{i'}X_{j'}]) = \\
    & = \sum_{i<j,\ i'<j'} J_{ij}^X J_{i'j'}^Y (w_{x_i}^{(k)} w_{x_j}^{(k)} w_{y_{i'}}^{(l)} w_{y_{j'}}^{(l)} - w_{x_i}^{(l)} w_{x_j}^{(l)} w_{y_{i'}}^{(k)} w_{y_{j'}}^{(k)}) [X_i X_j, Y_{i'}Y_{j'}] \ .
\end{eqnarray}
Using the triangle inequality, we can bound the norm of these commutators in the following way
\begin{equation}
    ||[H^{(k)}, H^{(l)}]|| \leq \sum_{i<j,\ i'<j'} J_{ij}^X J_{i'j'}^Y |w_{x_i}^{(k)} w_{x_j}^{(k)} w_{y_{i'}}^{(l)} w_{y_{j'}}^{(l)} - w_{x_i}^{(l)} w_{x_j}^{(l)} w_{y_{i'}}^{(k)} w^{y_{j'}}_k| \cdot ||[X_i X_j, Y_{i'}Y_{j'}]||
\end{equation}
and further, noting that $(w_{x_i}^{(k)} w_{x_j}^{(k)} w_{y_{i'}}^{(l)} w_{y_{j'}}^{(l)} - w_{x_i}^{(l)} w_{x_j}^{(l)} w_{y_{i'}}^{(k)} w_{y_{j'}}^{(k)}) \in \{-2, 0, 2\}$, we obtain
\begin{equation}
\label{eq:app_walsh_bound}
    ||[H^{(k)}, H^{(l)}]|| \leq 2 \sum_{i<j,\ i'<j'} J_{ij}^X J_{i'j'}^Y ||[X_i X_j, Y_{i'}Y_{j'}]||
\end{equation}
independently of $k$ and $l$. 
We have now obtained a bound on the norm of a commutator which does not take into account the structure of the specific Walsh sequence we are implementing, and therefore applies to the Trotter errors inherited from any Walsh sequence.

We now quantify the bound obtained from Eq.~\eqref{eq:app_walsh_bound} for the case of a one-dimensional long-range spin-exchange model, where
\begin{equation}
\label{eq:app_LR_coupling}
    J_{ij}^X = J_{ij}^Y = \frac{J}{|i-j|^{\alpha}} := J_{|i-j|}
\end{equation}
First, we re-arrange the sum
\begin{equation}
    \sum_{i<j,\ i'<j'} J^X_{ij} J^Y_{i'j'} ||[X_i X_j, Y_{i'} Y_{j'}]|| = \frac{1}{4} \sum_{i\neq j,\ i'\neq j'} J^X_{ij} J^Y_{i'j'} ||[X_i X_j, Y_{i'} Y_{j'}]|| \ . 
\end{equation}
Only the terms in the sum in which only one between $i$ or $j$ is equal to $i'$ or $j'$ will be non-zero, and in particular they will have spectral norm 2 (for the property $[X, Y] = 2iZ$, and $[X\otimes X,Y\otimes Y]=0$).
Since $J^X_{ij} = J^Y_{ji} = J_{|i-j|}$, we obtain 
\begin{equation}
\label{eq:app_err1}
    \sum_{i<j,\ i'<j'} J^X_{ij} J^Y_{i'j'} ||[X_i X_j, Y_i Y_j]|| = \sum_{i\neq j,  i\neq j'} J^X_{ij} J^Y_{ij'} - \sum_{i \neq j} J^X_{ij} J^Y_{ij} \leq \sum_{i\neq j,  i\neq j'} J_{|i-j|} J_{|i-j'|} \ .
\end{equation}
We can rewrite the obtained sum, and upper-bound it again
\begin{equation}
\label{eq:app_err2}
    \sum_{i\neq j,  i\neq j'} J_{|i-j|} J_{|i-j'|} = 4 \sum_{i<j,  i<j'} J_{|i-j|} J_{|i-j'|} = 4\sum_{r_1=1}^{N-1} \sum_{r_2=1}^{N-1} (N- \text{max}(r_1, r_2))J_{r_1}J_{r_2} \leq 4\sum_{r_1=1}^{N-1} (N- r_1)J_{r_1} \sum_{r_2=1}^{N-1} J_{r_2} 
\end{equation}
where we have used that $r_1$ and $r_2$ run from $1$ to $N - 1$, but at the same time $i \leq N - r_1$ and $i \leq N - r_2$. 
At this point, to simplify the computation of the sums, we use the general property of monotonously decreasing functions (coming from the definition of the Riemann integral)~\cite{rudin1976principles}
\begin{equation}
\label{eq:app_integral}
    \sum_{k = a}^b f(k) \leq f(a) + \int_a^b \text{d}x f(x) \ .
\end{equation}
The second factor in Eq.~\eqref{eq:app_err2}  admits the following integral upper bound
\begin{equation}
    \sum_{r_2=1}^{N-1} J_{r_2}  \leq J \left(1 + \int_1^{N-1} \text{d}r\ r^{-\alpha} \right) = J\frac{1}{1 - \alpha}((N-1)^{1-\alpha} - \alpha) 
\end{equation}
and similarly does the first factor 
\begin{equation}
\label{eq:app_integral2}
    \sum_{r_1=1}^{N-1} (N - r_1)J_{r_1}  \leq J \left( \frac{1}{(1 - \alpha)(2 - \alpha)}(N - 1)^{2 - \alpha} + \frac{1}{1 - \alpha}(N - 1)^{1 - \alpha} - \frac{\alpha}{1 - \alpha}N - \frac{(1-\alpha)}{2 - \alpha}\right) \ .
\end{equation}
Multiplying these two results yields then an upper bound to Eq.~\eqref{eq:app_err1}.
If one takes the large $N$ limit of the product it is possible to note that, depending on the value of $\alpha$, two possible leading terms arise.
For $\alpha>1$, one obtains
\begin{equation}
\label{eq:app_trotter_bound_g1}
    \sum_{i\neq j,  i\neq j'} J_{|i-j|} J_{|i-j'|} = 4J^2\left(\frac{\alpha}{\alpha - 1}\right)^2 N + o(N)  =: 2J^2a_{\alpha}N
\end{equation}
while for $\alpha < 1$
\begin{equation}
\label{eq:app_trotter_bound_l1}
    \sum_{i\neq j,  i\neq j'} J_{|i-j|} J_{|i-j'|} = 4J^2\frac{1}{(1 - \alpha)^2 (2 - \alpha)}N^{3 - 2\alpha} + o(N^{3 - 2\alpha})  =: 2J^2b_{\alpha}N^{3-2\alpha} \ .
\end{equation}
Therefore, any quantum simulation over a total time $T$ with $p = 1$ Walsh sequences with period $\tau$ will have a Trotter error bounded by
\begin{equation}
\label{eq:app_final_trotter}
\begin{cases}
    \mathcal{E} \leq \frac{\tau T}{2 n^2} \frac{n(n-1)}{2}\cdot 4J^2 a_{\alpha} N \leq a_{\alpha} N(JT)(J\tau) \ \text{if} \ \alpha > 1 \\
    \mathcal{E} \leq \frac{\tau T}{2 n^2} \frac{n(n-1)}{2}\cdot 4J^2 b_{\alpha} N^{3 - 2\alpha} \leq b_{\alpha} N^{3 - 2\alpha}(JT)(J\tau) \ \text{if} \ \alpha < 1
\end{cases}
\end{equation}
which implies the bound of Eq.~\eqref{eq:trotter_bound}, in the case $Q = 1$.

This result can be straightforwardly extended to cycles comprised of $Q$ Walsh sequences.
For each Walsh sequence, we have a different period $\tau_q$ and number of steps $n_q$.
Therefore, Eq.~\eqref{eq:app_general_lloyd} becomes
\begin{equation}
\label{eq:app_multi_lloyd}
    \mathcal{E}(\{\tau_q\}, T) = \frac{T}{2\sum_q \tau_q} \sum_{(q, k) < (q', l)} \frac{\tau_q \tau_{q'}}{n_q n_{q'}}||[H^{(q, k)},H^{(q', l)}]||
\end{equation}
where $H^{(q, k)}$ is the $k-$th toggling frame Hamiltonian of the $q-$th Walsh sequence, and $(q, k)<(q', l)$ means $k<l$ if $q=q'$, or $q<q'$ for arbitrary $k, l$.
The bound Eq.~\eqref{eq:app_walsh_bound} still holds, giving, for the previously considered example Eq.~\ref{eq:app_LR_coupling}
\begin{equation}
    ||[H^{(q, k)},H^{(q', l)}]|| \leq 2\sum_{i\neq j, i'\neq j'} J^X_{ij} J^Y_{ij} ||[X_i X_j, Y_{i'}Y_{j'}]|| \leq 2\sum_{i\neq j,  i\neq j'} J_{|i-j|} J_{|i-j'|} \ .
\end{equation}
Plugging this bound in Eq.~\eqref{eq:app_multi_lloyd} we obtain
\begin{equation}
    \mathcal{E}(\{\tau_q\}, T) = \frac{T}{\sum_q \tau_q} \left( \sum_q \sum_{k<l} \frac{\tau_q^2}{n_q^2} + \sum_{q<q'}\sum_{k, l} \frac{\tau_q \tau_{q'}}{n_q n_{q'}} \right)\sum_{i\neq j,  i\neq j'} J_{|i-j|} J_{|i-j'|} \le \frac{\left( \sum_q \tau_q \right) T}{2} \sum_{i\neq j,  i\neq j'} J_{|i-j|} J_{|i-j'|} \ .
\end{equation}
Now, plugging in the formula above the bounds of Eq.~\eqref{eq:app_trotter_bound_g1},~\eqref{eq:app_trotter_bound_l1}, we obtain expressions corresponding to Eq.~\eqref{eq:app_final_trotter} with the substitution $\tau \rightarrow \sum_q \tau_q$, i.e.
\begin{equation}
\label{eq:app_final_trotter_multi}
\begin{cases}
    \mathcal{E}(\{\tau_q\}, T) \leq a_{\alpha} N(JT)(J\sum_q\tau_q) \ \text{if} \ \alpha > 1 \\
    \mathcal{E}(\{\tau_q\}, T) \leq b_{\alpha} N^{3 - 2\alpha}(JT)(J\sum_q \tau_q) \ \text{if} \ \alpha < 1
\end{cases}
\end{equation}
which we report in Eq.~\eqref{eq:trotter_bound} in the main text.

\section{Approximation scheme for shorter Walsh sequences}
\label{app:errors2}

\begin{figure}[t]
    \centering
    \includegraphics[width=0.7\textwidth]{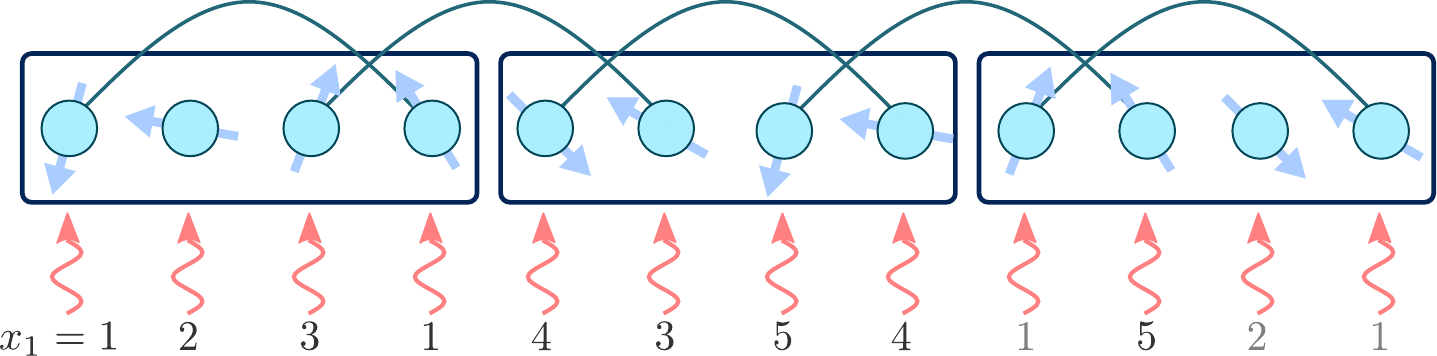}
    \caption{Example for the cut-off algorithm: the first assignment (Walsh indices dark grey) is for the first two blocks, and the second qubit of the third one. The second assignment covers the remaining qubits.}
    \label{fig:app_cutoff}
\end{figure}

In this section we investigate the unitary error due to imposing a cut-off $\Lambda_w$ to the Walsh indices.
We will focus on one-dimensional systems, even though we believe this approach can be extended to higher dimensional systems.
We consider the case in which the target Hamiltonian $\tilde{H}$ has finite range interactions, i.e. there exists a distance $r = O(1)$ such that $\tilde{J}_{ij}=0$ if $|i-j|>r$.
In general, the Walsh sequence realizing the dynamics of $\tilde{H}$ in AHT will have a length $n = O(N)$.
However, the length can be reduced by assuming that interactions in the resource Hamiltonian $H_R$ are negligible after a certain distance $\Lambda_r\geq r$ (for resource Hamiltonians with couplings Eq.~\eqref{eq:LR_coupling} it is the case if $\alpha$ is large).
In this case, we do not require using different Walsh indices to decouple two qubits that are separated by distance larger than $\Lambda_r$.
Formally, this means that Eq.~\eqref{eq:graph_equation} becomes
\begin{equation}
\label{eq:app_graph_cutoff}
    E(\mathcal{G}_1^X) = \{(i,j)\ |\ x_i=x_j\ \text{and}\ |i-j|<\Lambda_r \},\ E(\mathcal{G}_1^Y) = \{(i,j)\ |\ y_i=y_j\ \text{and}\ |i-j|<\Lambda_r \} \ .
\end{equation}
Fixing the distance cut-off $\Lambda_r$ provides us with a strategy to reduce the pulse sequence length (see Fig.~\ref{fig:app_cutoff}):
\begin{enumerate}
    \item order the values of Walsh indices $x_i$;
    \item divide the $N$ qubits in the subregions $A_l$, such that $[l\Lambda_r + 1, (l+1)\Lambda_r]\in A_l$, in such a way that links can only be present between $A_l$ and $A_{l+1}$;
    \item assign sequentially (the first $x_i$ to the 1st qubit) the Walsh indices to $A_1$ and $A_2$;
    \item assign sequentially the Walsh indices to $A_3$ starting from the beginning (i.e. from those in $A_1$), but excluding those already assigned in $A_2$
    \item repeat, where in any $A_l$ are excluded the Walsh indices in $A_{l-1}$.
\end{enumerate}
This protocol ensures that the shortest residual interaction acts on a distance $\Lambda_r + 1$.
Note that, if the region $A_l$ is not coupled with $A_{l-1}$, we can assign the Walsh indices there without constraints.
The highest Walsh index assigned in this way will be $\text{max}_i(x_i) \leq 2\Lambda_r$.
This can be proven from the fact that, given $X_l := \text{max}_{i \in A_l}(x_i)$, we have $X_{l} \leq \Lambda_r + X_{l-1} - X_{l-2}$ (together with $X_1\leq \Lambda_r$), as we can use the Walsh indices in $A_{l-2}$, but not those in $A_{l-1}$.
At the same time we can apply the condition, $X_{l-1} \leq \Lambda_r + X_{l-2}$, which further bounds the inequality above.
The two conditions together give $X_{l} \leq \Lambda_r + X_{l-1} - X_{l-2} \leq 2\Lambda_r$, as claimed.

Let us now address the effect of the residual interactions $|i-j|\ge \Lambda_r$, which we neglected in the first paragraph.
The Walsh sequence $(x_i, y_i)$ will not realize exactly (in AHT) the dynamics of the target Hamiltonian $\tilde{H}$, but instead, in a single Trotter step
\begin{equation}
    \tilde{U}(\tau) = e^{-i\tau(\tilde{H} + H_E)} + O(\tau^2) 
\end{equation}
where
\begin{equation}
    H_E = \sum_{i, j, |i-j|>\Lambda_r}(J^X_{ij}\delta_{x_i,x_j}X_i X_j + J^Y_{ij}\delta_{y_i,y_j} Y_i Y_j) \ .
\end{equation}
which consists in an additional contribution to the error $||(\tilde{U}(\tau))^{T/\tau} - e^{-iT\tilde{H}}||$.
We now provide an estimate for this contribution at the level of a single Trotter step.
First, we note that working in AHT allows us to regroup toggling-frame Hamiltonian arbitrarily, so 
\begin{equation}
    \tilde{U}(\tau) = e^{-i\tau\tilde{H} + H_E} + O(\tau^2) =  e^{-i\tau\tilde{H}}e^{-i\tau H_E} + O(\tau^2) = e^{-i\tau\tilde{H}}\left(1 + \sum_{l=1}^{\infty} \tau^l H_E^l \right) + O(\tau^2) \ .
\end{equation}
The norm of the error induced by $H_E$, beside Trotter errors, is then
\begin{equation}
    ||e^{-i\tau\tilde{H}}\left(1 + \sum_{l=1}^{\infty} \tau^l H_E^l \right) -  e^{-i\tau\tilde{H}}|| = ||e^{-i\tau\tilde{H}} \sum_{l=1}^{\infty} \tau^l H_E^l || \leq \sum_{l=1}^{\infty} \tau^l ||H_E||^l \ .
\end{equation}
To ensure that the error Hamiltonian $H_E$ does not hinder the simulation of the target Hamiltonian $\tilde{H}$, we would like the errors induced by $H_E$ to be comparable or smaller than the Trotter errors.
Therefore, we want them to be of order $O(\tau^2)$.
This can be obtained by fixing the norm of $H_E$ to be
\begin{equation}
\label{eq:app_precision_cutoff}
    ||H_E|| = \epsilon \tau
\end{equation}
with $\epsilon = O(1)$ to be a parameter tunable to make the error smaller or comparable to Trotter errors.
This translates in conditions on the cut-off distance $\Lambda_r$ and the exponent $\alpha$ of the interactions decay.
To obtain them, we evaluate a bound to $||H_E||$ for the one-dimensional case, of which we consider an example in Sec.~\ref{subsec:cutoff}.
From the triangle inequality
\begin{equation}
    ||H_E|| \leq \sum_{i=1}^{N-1} \sum_{r=\Lambda_r + 1}^{N-i} J_r ||X_i X_{i+r} + Y_i Y_{i+r}||  = \sum_{r=\Lambda_r + 1}^{N-1}(N - r)J_r \ .
\end{equation}
We now estimate the sum on the RHS using the integral bound Eq.~\eqref{eq:app_integral}, from which we obtain
\begin{equation}
    \sum_{r=\Lambda_r + 1}^{N-1}(N - r)J_r \leq \frac{J}{(1-\alpha)(2-\alpha)}(N-\alpha+1)(N-1)^{1-\alpha} - \frac{J}{1-\alpha}N(\Lambda_r +\alpha)(\Lambda_r +1)^{-\alpha} - \frac{J}{2-\alpha}(\Lambda_r + \alpha - 1)(\Lambda_r+1)^{1-\alpha} \ .
\end{equation}
By taking the large $N$ limit, we notice that $||H_E||$ has a different leading term depending on the value of $\alpha$.
For $\alpha < 1$ we have
\begin{equation}
    ||H_E|| \leq \frac{J}{(1-\alpha)(2-\alpha)}N^{2-\alpha} + O(N)
\end{equation}
which is a bound that does not depend on the cut-off $\Lambda_r$ and thus cannot guarantee the criterion Eq.~\eqref{eq:app_precision_cutoff}.
We deduce that, in this regime, imposing any cut-off to Walsh indices would lead to a dramatic error in the Hamiltonian engineering.
This can be also understood intuitively, by noticing that, for $\alpha<1$, even far away qubits are strongly interacting, therefore coupling them would lead to a much different dynamics than having them decoupled.
For $\alpha>1$ we have instead
\begin{equation}
    ||H_E|| \leq \frac{J}{(\alpha-1)}N(\Lambda_r +\alpha)(\Lambda_r +1)^{-\alpha} + O(1) \ .
\end{equation}
If we consider $\tau = O(N^{-1})$ (to have performance guarantees with respect to Trotter errors, see Eq.~\eqref{eq:app_final_trotter}), then for $\alpha>3$ this bound guarantees the existence of a value of $\Lambda_w = o(N)$ satisfying Eq.~\eqref{eq:app_precision_cutoff}.
In particular, $||H_E|| = \epsilon / N$ if $\Lambda_w = O((N/\epsilon)^{\frac{2}{\alpha - 1}})$.
Note that Trotter errors, from our numerical simulations, can have a weaker scaling with respect to $N$ for interesting quantities than what can be estimated by the unitary error, leading this approximation to work also for smaller $\alpha$.

Therefore, we conclude that, for $\alpha$ large enough, it is possible to set a cut-off $\Lambda_w = o(N)$ to reduce the length of Walsh sequences without introducing further errors.
In particular, for the one-dimensional case, we have provided analytical guarantee that this is the case for $\alpha>3$.
In Sec.~\ref{subsec:cutoff} we report a numerical example indicating that a finite $\Lambda_w$ will create errors comparable or smaller than Trotter errors even for $\alpha = 3$, but not for $\alpha < 1$.

\section{Rotation angle errors}
\label{app:rotation_angle_errors}

In this section, we derive the effect of rotation angle errors in the implementation of Walsh sequences, and derive conditions under which Walsh sequences are robust against this kind of errors.
In particular, we study their effect in AHT.
If the rotation angle errors are of order $O(\delta)$, we find that there is an error in AHT of order $O(\delta)$.
The robustness condition we find allows to reduce this AHT error to $O(\delta^2)$.

\subsection{Effect on the average Hamiltonian}

We model rotation angle errors by assuming that each $s_i\pi-$pulse in a certain direction (where $s_i = \pm 1$) becomes a $s_i(\pi + \delta_i)-$pulse instead, $\delta_i$ being a single-qubit error, i.e.
\begin{eqnarray}
    P^{(k)} = \bigotimes_i e^{-is_i(\pi + \delta_i)[O]_i/2}
\end{eqnarray}
where $[O]_i$ is the operator with respect to which we want to pulse on a given qubit.
This mimics the case in which the pulsing time is not well calibrated on the qubits.
The exact Walsh sequences are recovered for $\delta_i = 0$, and any choice of $s_i$.
The resource Hamiltonian considered is 
\begin{equation}
    H_R = \sum_{i<j} (J_{ij}^X X_i X_j + J_{ij}^Y Y_i Y_j) \ .
\end{equation}
To study the effect of the errors, we consider the first pulse of the interval $k$ to be $P^{(k)}$ as defined above, and the second one to be $(P^{(k)})^{-1}$.
The effect on local $X_i$ and $Y_i$ operators is
\begin{equation}
\begin{split}
    X_i^{(k)} := (p_i^{(k)})^{-1} X_i p_i^{(k)} &= \frac{1 + w_{x_i}^{(k)}}{2}X_i + \frac{1 - w_{x_i}^{(k)}}{2}(\cos(\pi + \delta_i)X_i + s_i\sin(\pi + \delta_i)\tilde{X}_i^{(k)}) = \\
    &= w_{x_i}^{(k)} X_i - \delta_i\frac{1 - w_{x_i}^{(k)}}{2}(s_i \tilde{X}_i^{(k)} - \frac{1}{2}\delta_i X_i) + O(\delta_i^3) \\
    Y_i^{(k)} := (p_i^{(k)})^{-1} Y_i p_i^{(k)} &= \frac{1 + w_{y_i}^{(k)}}{2}Y_i + \frac{1 - w_{y_i}^{(k)}}{2}(\cos(\pi + \delta_i)Y_i + s_i\sin(\pi + \delta_i)\tilde{Y}_i^{(k)}) = \\
    &= w_{y_i}^{(k)} Y_i - \delta_i\frac{1 - w_{y_i}^{(k)}}{2}(s_i \tilde{Y}_i^{(k)} - \frac{1}{2}\delta_i Y_i) + O(\delta_i^3)
\end{split}
\end{equation}
where
\begin{equation}
\begin{split}
    \tilde{X}_i^{(k)} = \left(\frac{1 - w_{y_i}^{(k)}}{2} Y_i - \frac{1 + w_{y_i}^{(k)}}{2}Z_i \right),\ \tilde{Y}_i^{(k)} = \left(\frac{1 + w_{x_i}^{(k)}}{2} Z_i - \frac{1 - w_{x_i}^{(k)}}{2}X_i \right) \ .
\end{split}
\end{equation}
The resulting toggling-frame Hamiltonian reads
\begin{equation}
\begin{split}
\label{eq:app_RA_tfH}
    H^{(k)} (\pmb{\delta}) = \sum_{i<j}\left(J^X_{ij}w^{(k)}_{x_i}w^{(k)}_{x_j}X_i X_j + J^Y_{ij}w^{(k)}_{y_i}w^{(k)}_{y_j} Y_i Y_j  \right) + H^{(k)}_{\text{RA}} = H^{(k)}(\pmb{\delta} = 0) + H^{(k)}_{\text{RA}} 
\end{split}
\end{equation}
where
\begin{equation}
\begin{split}
    H^{(k)}_{\text{RA}} = \sum_{O = X, Y}\sum_{i<j} J_{ij}^O \bigg( \left(s_i\delta_i \frac{1 - w^{(k)}_{o_i}}{2} w^{(k)}_{o_j} \tilde{O}_i^{(k)} O_j + s_j \delta_j w^{(k)}_{o_i}\frac{1 - w^{(k)}_{o_j}}{2} O_i \tilde{O}_j^{(k)} \right) + \\
    + s_i s_j \delta_i \delta_j \frac{1 - w^{(k)}_{o_i}}{2}\frac{1 - w^{(k)}_{o_j}}{2} \tilde{O}_i^{(k)}\tilde{O}_j^{(k)} +\frac{1}{2}\left(\delta_i^2 \frac{1 - w^{(k)}_{o_i}}{2} w^{(k)}_{o_j} + \delta_j^2 w^{(k)}_{o_i}\frac{1 - w^{(k)}_{o_j}}{2} \right)O_i O_j \bigg) + O(\delta^3) \ .
\end{split}
\end{equation}
This results tells us that the rotation angle errors will ultimately lead to errors in the average Hamiltonian of order $O(\delta)$.
In the next subsection we derive a condition to remove the $O(\delta)$ error from the average Hamiltonian.

\subsection{Robustness conditions}

We now show how it is possible to use the sign functions $s_i$ to remove the order $O(\delta)$ in the rotation angle error term of the average Hamiltonian.
We start by labelling a single Walsh sequence (or a cycle of multiple Walsh sequences) with an index $l$.
This means that the cycles $\tilde{U}(\tau)$ (which can be made by multiple Walsh sequence) will be now labelled as $\tilde{U}^{(l)}(\tau)$, such that
\begin{equation}
    (\tilde{U}(\tau))^{T/\tau} \rightarrow \prod_{l=1}^{T/\tau}\tilde{U}^{(l)}(\tau) \ .
\end{equation}
For later convenience, we choose the index $l$ to have a periodicity $L$, i.e.
\begin{equation}
    \tilde{U}^{(l+L)}(\tau) = \tilde{U}^{(l)}(\tau) \ .
\end{equation}
This is also known as second averaging~\cite{choi_robust_2020}.
We now consider the sign functions $s_i$ to be dependent on this new index $l$, i.e. $s_i \rightarrow s_i^l$.
Note that this does not change the results for Walsh sequences without pulse errors.
Let us further choose $s_i^l = w^{(l)}_{e_i}$.
We then introduce the second index in the toggling-frame Hamiltonians $H^{(k, l)}$, and the doubly averaged Hamiltonian
\begin{equation}
    \tilde{H}'_{\text{RA}} = \frac{1}{L}\sum_{l=1}^L\tilde{H}^{(l)} = \frac{1}{nL}\sum_{l=1}^L \sum_{k=1}^n H^{(k, l)} 
\end{equation}
where $L$ is the period of the longest Walsh function $w_{e_i}$.
Assuming $x_i, y_i \neq 0$, performing the double average over the toggling-frame Hamiltonians in Eq.~\eqref{eq:app_RA_tfH}, we obtain 
\begin{equation}
\begin{split}
    \tilde{H}'_{\text{RA}} = \sum_{O = X, Y}\sum_{i<j} J_{ij}^O \Bigg(\delta_{o_i, o_j}\left( 1 - \frac{\delta_i^2 + \delta_j^2}{4} \right)O_i O_j + \left(\delta_{0, e_i}\delta_i \mathcal{O}_i^j O_j + \delta_{0, e_j} \delta_j O_i \mathcal{O}_j^i \right) +  \delta_{e_i, e_j} \delta_i \delta_j \mathcal{C}_{ij} \Bigg) + O(\delta^3) 
\end{split}
\end{equation}
where the sum over $O = X, Y$ runs also over $o = x, y$ for the Walsh indices, and
\begin{equation}
\begin{split}
    \mathcal{O}_i^j =& \frac{1}{n}\sum_{k=1}^n \frac{1 - w^{(k)}_{o_i}}{2}w^{(k)}_{o_j}\tilde{O}^{(k)}_i \\
    \mathcal{C}_{ij} =& \frac{1}{n}\sum_{k=1}^n \frac{1 - w^{(k)}_{o_i}}{2}\frac{1 - w^{(k)}_{o_j}}{2}\tilde{O}^{(k)}_i\tilde{O}^{(k)}_j \ .
\end{split}
\end{equation}
From the equations above, it can be noticed that choosing $e_i \neq 0$ $\forall i$ removes exactly the undesired term of $O(\delta)$.
We will refer to this condition as the robustness condition against rotation angle errors.
The easiest choice to implement the robustness condition is $e_i = 1$ $\forall i$, which amounts to performing alternatively $\pi-$ and $-\pi-$ pulses in consecutive Walsh sequences.
Another choice is instead $e_i = i$, which furthermore removes the term $\sim \delta_i \delta_j$ from the doubly averaged Hamiltonian.
Upon choosing $e_i = i$, the doubly averaged Hamiltonian reads
\begin{equation}
\begin{split}
    \tilde{H}'_{\text{RA}} = \sum_{i<j} \left( 1 - \frac{\delta_i^2 + \delta_j^2}{4} \right)\left( J_{ij}^X\delta_{x_i, x_j}X_i X_j + J_{ij}^Y\delta_{y_i, y_j}Y_i Y_j \right) + O(\delta^3) 
\end{split}
\end{equation}
which amounts to the target Hamiltonian, plus a $O(\delta^2)$ inhomogeneous renormalization of the interactions.

\begin{figure}[t]
    \includegraphics[width=0.9\textwidth]{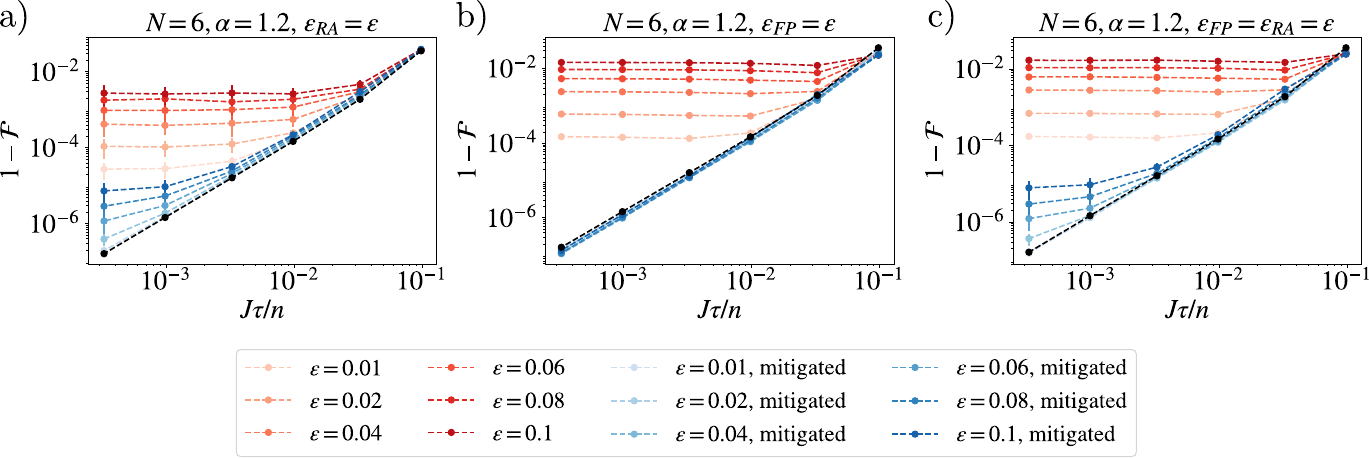}
    \caption{\textit{Ising dynamics with robust Walsh sequences} $-$ We study the effects of pulse imperfections in the cluster state preparation protocol from Ising dynamics described in Sec.~\ref{sec:ising_chain}. We report numerical data for a chain of $N = 6$ qubits, with $\alpha = 1.2$. 
    a) The red lines represent the fidelity error $1 - |\langle \psi_{\text{sim}}|\psi_{\text{cluster}}\rangle|$ in presence of a systematic rotation angle error $\delta_i$ on the $i-$th qubit, $\delta_i$ sampled uniformly in $[-\epsilon_{\text{RA}}J, \epsilon_{\text{RA}}J]$ (datas are averaged over 64 samples of $\{\delta_i\}$).
    We observe this error to reduce of order of magnitude by implementing the robustness condition (blue lines), while being still above the Trotter error for perfect pulses (black line).
    b) We observe the increase in the fidelity error also for finite pulses of duration $t_p = \frac{\tau}{2n}\epsilon_{\text{FP}} $ (red lines). Implementing the robustness condition, the fidelity error collapses on the Trotter error for perfect pulses (black line).
    In c) we plot the total effect of both sources of error (red lines), and its reduction upon implementing the robustness conditions.
    }
    \label{fig:app_robustness}
\end{figure}

A numerical example of the implementation of this robustness condition is shown in Fig.~\ref{fig:app_robustness}.
In particular, we consider the cluster state preparation protocol via Ising dynamics (presented in Sec.~\ref{sec:ising_chain}) for $N=6$ qubits, and $\alpha = 1.2$.
We consider a uniform sampling of each $\delta_i$ from the interval $[-\epsilon_{\text{RA}}J, \epsilon_{\text{RA}}J]$, for different values of $\epsilon_{RA}$.
We quantify the errors by plotting the fidelity error with respect to the cluster state, averaged over $64$ $\delta_i$ samples, both for the case in which the robustness condition is implemented (blue lines) and not implemented (red lines), and compare it with the pure Trotter errors.
It can be observed that implementing the robustness condition reduces the fidelity error by orders of magnitude.
While the fidelity error for robust Walsh sequences still saturates for small values of $\tau$ (due to the residual $O(\delta^2)$ term), this saturation happens for much smaller values of $\tau$, leading to a relevant improvement of the protocol. 

\section{Finite pulse duration errors}
\label{app:finite_pulse_duration_errors}

In this section we consider the effect of pulses of finite duration with respect to the ideal case, in which the pulses are instantaneous, and their dynamics is decoupled from the Hamiltonian dynamics given by interactions.
We first derive the effect on the average Hamiltonian, which are of order $O(t_p/\tau)$, and then we show a robustness condition which completely removes the errors in AHT, upon a slight modification of the simulation procedure.

\subsection{Average Hamiltonian theory for finite pulse duration}

Let consider an interval $k$ of a Walsh sequence, where we want to apply a pulse $P^{(k)}$ at the beginning, and another pulse $(P^{(k)})^{-1}$ at the end.
We assume that the pulses have a square-wave form in time.
In order to apply a $s_i\pi-$pulse in the $[O]_i$ direction, the physical implementation is done by means of the Hamiltonian
\begin{equation}
    H_p = \frac{\pi}{2t_p}\sum_i s_i [O]_i
\end{equation}
where 
\begin{equation}
    P^{(k)} = e^{-it_p H_p},\ (P^{(k)})^{-1} = e^{it_p H_p}
\end{equation}
and $t_p$ is the time duration of the pulse.
In general, during the application of a pulse, the resource Hamiltonian will still make the qubits of the underlying system interact.
Therefore, the evolution during an interval $k$ of a Walsh sequence will be
\begin{equation}
    U_{\frac{\tau}{n}}^{(k)} = e^{-it_p(H_R - H_p)}e^{-i(\tau/n - 2t_p)H_R}e^{-it_p(H_R + H_p)} \ .
\end{equation}
In the case of instantaneous pulses ($t_p=0$), considered throughout the text, the evolution above becomes
\begin{equation}
    U_{\frac{\tau}{n}}^{(k)} = (P^{(k)})^{-1}e^{-i(\tau/n)H_R}P^{(k)} = e^{-i(\tau/n)H^{(k)}} 
\end{equation}
i.e. the pulses and the interacting evolution are decoupled, as expected.
To analyze the relation between the instantaneous pulses case and the finite pulses case, we put ourselves in the interaction picture with respect to the pulsing Hamiltonians $H_p$, obtaining~\cite{Haeberlen68}
\begin{equation}
\begin{split}
    e^{-it_p(H_R \pm H_p)} = e^{\mp it_p H_p}\mathcal{U}_{\pm}(t_p) &= (P^{(k)})^{\pm 1}\mathcal{U}_{\pm}(t_p) \\ 
    \frac{d\mathcal{U}_{\pm}(t)}{dt} =  -i(e^{\pm it H_{p}} H_R e^{ \mp it H_{p}})\mathcal{U}_{\pm}(t) &= -i((Q^{(k)}(t))^{\mp 1} H_R (Q^{(k)}(t))^{\pm 1}\mathcal{U}_{\pm}(t) 
\end{split}
\end{equation}
where we have defined the partially applied pulse
\begin{equation}
\begin{split}
    Q^{(k)}(t) = e^{-itH_p} = \bigotimes_i q^{(k)}_i(t) \ .
\end{split}
\end{equation}
Hence, we have
\begin{equation}
\begin{split}
    U^{(k)}_{\frac{\tau}{n}}  =  (P^{(k)})^{-1}\mathcal{U}_-(t_p) e^{-i(\tau/n - 2t_p))H_R} P^{(k)} \mathcal{U}_+(t_p)  \ .
\end{split}
\end{equation}
Defining the toggling-frame Hamiltonian $H^{(k)} = (P^{(k)})^{-1}H_R P^{(k)}$, we have
\begin{equation}
    U^{(k)}_{\frac{\tau}{n}} = \mathcal{U}'_-(t_p) e^{-i(\tau/n - 2t_p))H^{(k)}}\mathcal{U}_+(t_p)
\end{equation}
where 
\begin{equation}
    \frac{d\mathcal{U}'_-(t)}{dt} = (P^{(k)})^{-1} \frac{d\mathcal{U}_-(t)}{dt} P^{(k)}= -i(Q^{(k)}(t) H^{(k)}  (Q^{(k)}(t))^{-1})\mathcal{U}_-'(t) \ .
\end{equation}
We can then define the instantaneous toggling-frame Hamiltonian for the interval $k$ as follows:
\begin{equation}
\begin{split}
    H^{(k)}(t) = (Q^{(k)}(t))^{-1} H_R Q^{(k)}(t)&,\ t<t_p \\
    H^{(k)}(t) = (P^{(k, 1)})^{-1} H_R P^{(k, 1)} = H^{(k)}&,\ t_p<t<\frac{\tau}{n} - t_p \\
    H^{(k)}(t) = Q^{(k)}(t) H^{(k)} (Q^{(k)}(t))^{-1} &,\ \frac{\tau}{n} - t_p < t < \frac{\tau}{n} \ .
\end{split}
\end{equation}
The average Hamiltonian over the interval $k$ is then~\cite{magnus1954exponential}
\begin{equation}
    H^{(k)}_{\text{av}} = \frac{n}{\tau}\int_0^{\tau/n} dt\ H^{(k)}(t) 
\end{equation}
and over the sequence is
\begin{eqnarray}
    \tilde{H} = \frac{1}{n} \sum_{k=1}^n H^{(k)}_{\text{av}} = \frac{1}{n} \sum_{k=1}^n \frac{n}{\tau} \int_0^{\tau/n} dt\ H^{(k)}(t) \ .
\end{eqnarray}
It is already possible to note that the average Hamiltonian over the interval $k$ reads
\begin{equation}
\begin{split}
    H^{(k)}_{\text{av}} =& \left(1 - \frac{2nt_p}{\tau} \right)H^{(k)} + \frac{n}{\tau} \int_0^{t_p}dt\ \left((Q^{(k)}(t))^{-1} H_R Q^{(k)}(t) + Q^{(k)}(t) H^{(k)} (Q^{(k)}(t))^{-1} \right) = \\
    =& \left(1 - \frac{2nt_p}{\tau} \right)H^{(k)} + \frac{2n}{\tau}\int_0^{t_p}dt\ (Q^{(k)}(t))^{-1} H_R Q^{(k)}(t)
\end{split}
\end{equation}
from which we can define the dimensionless error parameter
\begin{equation}
    \epsilon_{\text{FP}} := \frac{2nt_p}{\tau}
\end{equation}
which amounts to the fraction of time of the interval $k$ during which pulses are applied.
Finally, we define the average Hamiltonian over the whole sequence for finite pulses
\begin{equation}
\label{eq:app_avg_H_FP_general}
    \tilde{H}_{\text{FP}} = \left(1 - \epsilon_{\text{FP}} \right)\tilde{H} + \epsilon_{\text{FP}}\frac{1}{n}\sum_{k=1}^n\frac{1}{t_p}\int_0^{t_p}dt\ (Q^{(k)}(t))^{-1} H_R Q^{(k)}(t)
\end{equation}
where $\tilde{H}$ is the target Hamiltonian.
We conclude then that we only need to compute the second term on the right-hand side to estimate the finite pulse duration effects.

\subsection{Average Hamiltonian with finite pulse duration effects}

We now apply the result above to the Walsh sequence protocol considered in the text.
We consider the long-range $XY$ model as a resource Hamiltonian
\begin{eqnarray}
    H_R = \sum_{i<j} (J_{ij}^X X_i X_j + J_{ij}^Y Y_i Y_j) \ .
\end{eqnarray}
In order to compute the instantaneous toggling-frame Hamiltonian in an interval $k$, we need to keep track of the evolution of the instantaneous toggling-frame $X$ and $Y$ operators, which evolve during the first pulse ($0<t<t_p$) as
\begin{equation}
    X^{(k)}_i (t) :=(q_i^{(k)}(t))^{-1} X_i q_i^{(k)}(t) = \frac{1 + w^{(k)}_{x_i}}{2} X_i + \frac{1 - w^{(k)}_{x_i}}{2}\left( X_i \cos(\pi (t / t_p ) ) + s_i\tilde{X}^{(k)}_i \sin(\pi (t / t_p )) \right)  
\end{equation}
and 
\begin{eqnarray}
    Y^{(k)}_i (t) :=(q_i^{(k)}(t))^{-1} Y_i q_i^{(k)}(t) = \frac{1 + w^{(k)}_{y_i}}{2} Y_i + \frac{1 - w^{(k)}_{y_i}}{2}\left( Y_i \cos(\pi (t / t_p ) ) + s_i\tilde{Y}^{(k)}_i \sin(\pi (t / t_p )) \right) 
\end{eqnarray}
where
\begin{equation}
\begin{split}
    \tilde{X}_i^{(k)} = \left(\frac{1 - w^{(k)}_{y_i}}{2} Y_i - \frac{1 + w^{(k)}_{y_i}}{2}Z_i \right),\ \tilde{Y}_i^{(k)} = \left(\frac{1 + w^{(k)}_{x_i}}{2} Z_i - \frac{1 - w^{(k)}_{x_i}}{2}X_i \right) \ .
\end{split}
\end{equation}
We can now compute the integral in Eq.~\eqref{eq:app_avg_H_FP_general} as 
\begin{equation}
\begin{split}
    \frac{1}{t_p}\int_0^{t_p} dt\ (Q^{(k)}(t))^{-1} H_R Q^{(k)}(t) = \frac{1}{\pi}\int_0^{\pi} d\theta\ H^{k}(\theta) = \frac{1}{\pi} \sum_{O = X, Y} \sum_{i<j} J_{ij}^O \int_0^{\pi} d\theta\ O_i^{(k)}(\theta)O_j^{(k)}(\theta)
\end{split}
\end{equation}
where
\begin{equation}
\begin{split}
    O^{(k)}_i (\theta) = \frac{1 + w^{(k)}_{o_i}}{2} O_i + \frac{1 - w^{(k)}_{o_i}}{2}\left( O_i \cos\theta  + s_i\tilde{O}^{(k)}_i \sin\theta \right) \ .
\end{split}
\end{equation}
Making use of the following trigonometric integrals
\begin{equation}
\begin{split}
    \int_0^{\pi} d\theta = \pi,\ 
    \int_0^{\pi} d\theta \cos\theta = 0,\ 
    \int_0^{\pi} d\theta \sin\theta = 2, \\ 
    \int_0^{\pi} d\theta \sin\theta \cos\theta = 0,\ 
    \int_0^{\pi} d\theta \cos^2\theta =  \frac{\pi}{2},\ 
    \int_0^{\pi} d\theta \sin^2\theta =  \frac{\pi}{2} \ .
\end{split}
\end{equation}
we obtain
\begin{equation}
\begin{split}
    \frac{1}{\pi}\int_0^{\pi} d\theta\ O^{(k)}_i(\theta) O^{(k)}_j(\theta) =  \frac{1}{8}\left( 3 - w^{(k)}_{o_i} - w^{(k)}_{o_j} + 3 w^{(k)}_{o_i}w^{(k)}_{o_j} \right) O_i O_j + \frac{1}{2}\frac{1 - w^{(k)}_{o_i}}{2} \frac{1 - w^{(k)}_{o_j}}{2} s_i s_j \tilde{O}_i^{(k)} \tilde{O}_j^{(k)} + \\
    + \frac{2}{\pi}\left(s_i \frac{1 - w^{(k)}_{o_i}}{2}\frac{1 + w^{(k)}_{o_j}}{2}\tilde{O}_i^{(k)} O_j + s_j \frac{1 + w^{(k)}_{o_i}}{2}\frac{1 - w^{(k)}_{o_j}}{2} O_i \tilde{O}_j^{(k)} \right) \ .
\end{split}
\end{equation}

Performing the sum over the intervals $k$, we obtain that finite pulse duration effects contribute to the average Hamiltonian in the following way
\begin{equation}
\begin{split}
    \frac{1}{n}\sum_{k=1}^n \frac{1}{t_p} \int_0^{t_p}dt\ (Q^{(k)}(t))^{-1} H_R Q^{(k)}(t) &= \sum_{O = X, Y}\sum_{i<j}J^O_{ij}\left( \frac{3}{8}(1 + \delta_{o_i, o_j})O_i O_j +  \frac{s_i s_j}{2}\mathcal{C}_{ij} + \frac{2}{\pi}\left( s_i {\mathcal{O}'}_i^j O_j + s_j O_i {\mathcal{O}'}_j^i \right)\right) = \\
    &= \frac{3}{8}\tilde{H} + \frac{3}{8}H_R + \tilde{H}_{\text{FP,err}}^{(1)} + \tilde{H}_{\text{FP,err}}^{(2)}
\end{split}
\end{equation}
where 
\begin{equation}
\begin{split}
    \mathcal{C}_{ij} =& \frac{1}{n}\sum_{k=1}^n \frac{1 - w^{(k)}_{o_i}}{2}\frac{1 - w^{(k)}_{o_j}}{2}\tilde{O}^{(k)}_i\tilde{O}^{(k)}_j \\
    {\mathcal{O}'}_i^j =& \frac{1}{n}\sum_{k=1}^n \frac{1 - w^{(k)}_{o_i}}{2}\frac{1 + w^{(k)}_{o_j}}{2}\tilde{O}^{(k)}_i 
\end{split}
\end{equation}
and
\begin{equation}
\begin{split}
    \tilde{H}^{(1)}_{\text{FP, err}} =& \frac{1}{2}\sum_{O = X, Y}\sum_{i<j}J_{ij}^O s_i s_j\mathcal{C}_{ij}  \\
    \tilde{H}^{(2)}_{\text{FP, err}} =& \frac{2}{\pi}\sum_{O = X, Y}\sum_{i<j}J_{ij}^O\left( s_i {\mathcal{O}'}_i^j O_j + s_j O_i {\mathcal{O}'}_j^i \right) \ .
\end{split}
\end{equation}
Summing all the contribution to the average Hamiltonian, we obtain
\begin{equation}
\begin{split}
    \tilde{H}_{\text{FP}} =\left(1 - \frac{5 \epsilon_{\text{FP}}}{8} \right)\tilde{H} + \frac{3\epsilon_{\text{FP}}}{8}H_R + \epsilon_{\text{FP}}(\tilde{H}^{(1)}_{\text{FP, err}} + \tilde{H}^{(2)}_{\text{FP, err}}) \ .
\end{split}
\end{equation}
In the next subsection, we show how to remove the last two terms in AHT using the same second averaging as done with rotation angle errors, and to engineer $\tilde{H}$ up to Trotter errors even with finite pulse duration effects.

\subsection{Robustness conditions}

As already done with rotation angle errors, we now show that it is possible to exploit the sign functions $s_i$ to mitigate the finite pulse duration errors.
We choose the sign functions to change from a cycle to the other as $s_i^l = w^{(l)}_{e_i}$.
Averaging the Hamiltonian errors over $L$ cycles, $L$ being the periodicity of the Walsh functions $w_{e_i}$ with largest Walsh index, gives
\begin{equation}
\begin{split}
    \frac{1}{L}\sum_{l = 1}^L \tilde{H}^{(1, l)}_{\text{FP, err}} =& \frac{1}{2}\sum_{O = X, Y}\sum_{i<j}J_{ij}^O \delta_{e_i, e_j}\mathcal{C}_{ij}  \\
    \frac{1}{L}\sum_{l = 1}^L \tilde{H}^{(2, l)}_{\text{FP, err}} =& \frac{2}{\pi}\sum_{O = X, Y}\sum_{i<j}J_{ij}^O\left( \delta_{0, e_i} {\mathcal{O}'}_i^j O_j + \delta_{0, e_j} O_i {\mathcal{O}'}_j^i \right)
\end{split}
\end{equation}
Then, choosing $e_i \neq e_j$ $\forall i, j$ and $e_i\neq 0$ $\forall i$, the doubly averaged Hamiltonian becomes
\begin{equation}
\label{eq:app_corr_avg_H_FP}
\begin{split}
    \tilde{H}'_{\text{FP}} =\left(1 - \frac{5 \epsilon_{\text{FP}}}{8} \right)\tilde{H} + \frac{3\epsilon_{\text{FP}}}{8}H_R \ .
\end{split}
\end{equation}
To recover $\tilde{H}$ in AHT, we first take care of the second term, which it is possible to get rid of by simulating the interval $k=1$ (for which $H^{(1)} = H_R$) for a reduced time.
This can be seen by expanding in Eq.~\eqref{eq:app_corr_avg_H_FP} the average over pulses with respect to the first term
\begin{equation}
\begin{split}
    \left(\tilde{H}'_{\text{FP}} - \frac{1}{n}H_R\right) + \frac{1}{\tau}\left(\frac{\tau}{n} - \frac{3nt_p}{4} \right) H_R =\left(1 - \frac{5 \epsilon_{\text{FP}}}{8} \right)\tilde{H} 
\end{split}
\end{equation}
resulting in the reduced time $\frac{\tau}{n} - \frac{3nt_p}{4}$ for the interval $k=1$.
Finally, the factor $(1 - \frac{5 \epsilon_{\text{FP}}}{8})$ in front of $\tilde{H}$ just leads to the renormalization of the total simulation time $T$, which can be taken care of by increasing it as $T \rightarrow T\left(1 - \frac{5 \epsilon_{\text{FP}}}{8}\right)^{-1}$.
We note that these robustness conditions exactly remove finite pulse duration errors in AHT.

We benchmark numerically the effect of these robustness conditions with the cluster state preparation protocol via Ising dynamics for $N=6$ qubits.
The numerical results for the fidelity error are plotted in Fig.~\ref{fig:app_robustness}, where we observe that the fidelity error, being in AHT when the robustness conditions are not respected, collapse to the Trotter error when they are met instead.

As a last note, we highlight how the robustness conditions to mitigate finite pulse duration errors ($e_i \neq e_j$, $e_i\neq 0$) also imply the robustness against rotation angle errors.
This means that it is possible to mitigate both errors at the same time, without additional requirements than those asked to mitigate finite pulse duration errors.
While this mitigates exactly the individual contributions of these two sources of errors, there might be additional errors due to their mutual contribution, which are not considered here.
A numerical example for the combined mitigation of finite pulse duration and rotation angle errors in plot in Figs.~\ref{fig:cutoff},~\ref{fig:app_robustness}, which shows a reduction of the fidelity error in the cluster state preparation protocol of orders of magnitude.

\section{Dynamical decoupling with Walsh functions}
\label{app:dynamical_decoupling}

We derive the conditions for dynamical decoupling with Walsh functions from external disorder fields.
As done in the main text, we consider a resource Hamiltonian which also has space-dependent single-qubit fields
\begin{equation}
    H_R' = H_R + H_{\text{ext}}
\end{equation}
with
\begin{equation}
    H_{\text{ext}} = \sum_i (h^x_i X_i + h^y_i Y_i + h^z_i Z_i) \ .
\end{equation}
We now compute the effect of a Walsh sequence on the Hamiltonian term $H_{\text{ext}}$.
Applying a Walsh sequence, we note that the $X$ disorder term would become
\begin{equation}
    \tilde{H}_{\text{ext}}^X = \frac{1}{n}\sum_{k=1}^n \sum_i  h^x_i w^{(k)}_{x_i} X_i = \sum_i \left(\frac{1}{n}\sum_{k=1}^n w^{(k)}_{x_i} \right) h^X_i X_i = \sum_i \left(\frac{1}{n}\sum_{k=1}^n w_0^{(k)} w^{(k)}_{x_i} \right) h^X_i X_i = \sum_i \delta_{0, x_i} h^X_i X_i
\end{equation}
since $w^0_k = 1\ \forall k$, and applying the orthonormality condition for Walsh functions.
The same reasoning can be applied to the $Y$ part.
Finally, we note that the sign of the $Z$ operator is constrained by the signs of $X$ and $Y$, since
\begin{equation}
    p^{(k)}_i Z (p^{(k)}_i)^{-1} = i p^{(k)}_i X (p^{(k)}_i)^{-1} p^{(k)}_i Y (p^{(k)}_i)^{-1} = i (w^{(k)}_{x_i}X_i) (w^{(k)}_{y_i}Y_i)  = w^{(k)}_{x_i} w^{(k)}_{y_i} Z_i
\end{equation}
which implies
\begin{equation}
    \tilde{H}_{\text{ext}}^Z = \sum_i \left(\frac{1}{n}\sum_{k=1}^n w^{(k)}_{x_i} w^{(k)}_{y_i} \right) h^Z_i Z_i = \sum_i \delta_{x_i, y_i} h^Z_i Z_i 
\end{equation}
applying the orthonormality conditions for Walsh functions.

\section{Derivation of the quantum circuit for the surface code}
\label{app:surface_proof}

We show how to express the quantum circuit for the stabilizer measurement Fig.~\ref{fig:surface_code}a) in terms of Ising interactions and single-qubit rotations. 
Our starting point is the expression of the gate $C_{XX}=HC_{NOT}H$ 
\begin{equation}
C_{XX} = H(
\ket{0}\bra{0}\otimes \mathbf{1}
+\ket{1}\bra{1}\otimes X)H=
\ket{0}_X\bra{0}\otimes \mathbf{1}
+\ket{1}_X\bra{1}\otimes X
\end{equation}
with $\ket{a}_X=H\ket{a}$ the eigenstates of the $X$ operator.
This can be mapped to Ising interactions as follows
\begin{align}
C_{XX} 
&=\exp(i \pi (\ket{1}_X\bra{1}\otimes \ket{1}_X\bra{1}))
\nonumber \\
&= \exp(i (\pi/4) (1-X)\otimes (1-X))
\nonumber \\
&= R_{XX}^{\pi/2}(R_X^{\pi/2}\otimes R_X^{\pi/2})
\end{align}
with $R_{XX}^{\pi/2}=\exp(i X\otimes X \pi / 4)$, $R_X^{\pi/2}=\exp(-i X \pi / 4)$. Note that the last equality holds up to an irrelevant global phase. 
This leads directly to the circuits shown in Fig.~\ref{fig:surface_code}.

\end{document}